\documentclass[a4paper,fleqn,usenatbib]{mnras}
\pdfoutput=1
\usepackage{mathptmx}

\usepackage[T1]{fontenc}
\usepackage{ae,aecompl}
\usepackage{psfrag}
\usepackage{epstopdf}
\usepackage{cleveref}
\usepackage{units}
\usepackage{float}

\usepackage{graphicx}	
\usepackage{amsmath}	
\usepackage{amssymb}	
\usepackage{amsfonts}
\usepackage{color,soul}
\usepackage{verbatim}
\usepackage{tabu}
\usepackage{tikz}
\title[Red Misfits in the SDSS]{Red Misfits in the Sloan Digital Sky Survey: Properties of Star-Forming Red Galaxies}

\author[F. Evans et al.]{
Fraser A. Evans,\thanks{E-mail: evansfa@mcmaster.ca}
Laura C. Parker,
Ian D. Roberts 
\\
Department of Physics and Astronomy, McMaster University, 1280 Main Street West, Hamilton, ON, L8S 4L8, Canada\\
}

\date{Accepted XXX. Received YYY; in original form ZZZ}

\pubyear{2018}

\begin{document}
\label{firstpage}
\pagerange{\pageref{firstpage}--\pageref{lastpage}}
\maketitle

\begin{abstract}
  We study Red Misfits, a population of red, star-forming galaxies in the local Universe. We classify galaxies based on inclination-corrected optical colours and specific star formation rates derived from the Sloan Digital Sky Survey Data Release 7. Although the majority of blue galaxies are star-forming and most red galaxies exhibit little to no ongoing star formation, a small but significant population of galaxies ($\sim$11 per cent at all stellar masses) are classified as red in colour yet actively star-forming. We explore a number of properties of these galaxies and demonstrate that \textcolor{black}{Red Misfits are not simply dusty or highly-inclined blue cloud galaxies or quiescent red galaxies with poorly-constrained star formation.} The proportion of Red Misfits is nearly independent of environment and this population exhibits both intermediate morphologies and an enhanced likelihood of hosting an AGN. We conclude that Red Misfits are a transition population, gradually quenching on their way to the red sequence and this quenching is dominated by internal processes rather than environmentally-driven processes. \textcolor{black}{We discuss the connection between Red Misfits and other transition galaxy populations, namely S0's, red spirals and green valley galaxies.}
  
\end{abstract}



\section{Introduction}

Large galaxy redshift surveys such as the Sloan Digital Sky Survey \citep[SDSS;][]{York2000} have demonstrated that the distributions of \textcolor{black}{optical} colour and star formation rate (SFR) for galaxies in the local Universe are bimodal  \citep[e.g.][]{Strateva2001,Blanton2003, Kauffmann2003a, Baldry2004, Balogh2004b, Balogh2004a}. Galaxies are almost all \textcolor{black}{either red with early-type}  morphologies and little-to-no ongoing star formation, or blue and late-type with substantial ongoing star formation. The origin and maintenance of this bimodality is an important open question in galaxy formation and evolution.

Exploring how colour, morphology and star formation rate correlate with other galaxy properties \textcolor{black}{may help in understanding the origin and maintenance of the bimodality in galaxy populations}. The strong correlation between these properties and stellar mass \citep[e.g.][]{Kauffmann2003b, Noeske2007a,Noeske2007b, Wetzel2012} suggests that a galaxy's stellar mass is the main parameter upon which its other properties depend \citep[e.g.][]{Peng2010,Woo2013}. 

In addition, it has long been known that galaxy properties differ across environments: galaxy demographics in rich environments such as the cores of galaxy clusters are dominated by red, early-type galaxies with negligible star formation, while blue, star-forming, late-type galaxies are most commonly found in the low density field \citep[e.g.][]{Oemler1974, Dressler1980, Lewis2002, Gomez2003, Balogh2004a}. These correlations could simply be due to more massive galaxies residing in denser environments \citep[e.g.][]{Hogg2003}, however recent studies show that these trends persist even at fixed stellar mass \citep{Kauffmann2004, Blanton2005a, Skibba2009, McGee2011}. Not limited to the local Universe, this environmental dependence has been observed out to at least z$\approx$1 \citep{Cooper2007,Peng2010}. Together, these results motivate a complex picture of galaxy evolution wherein internal mechanisms scaling with stellar mass and mechanisms scaling with environment work in tandem to transform blue, star-forming galaxies to red and quiescent ones \citep[e.g.][]{vandenBosch2008a, Peng2010, Peng2012, Woo2013, Smethurst2017}. This evolutionary picture is supported by the fact that the total amount of stellar mass in red sequence galaxies has roughly doubled since z=1 \citep[e.g.][]{Bell2004, Faber2007, Moutard2016}.

The exact balance of environmental and internal mechanisms that dominate this transition is unclear and many mechanisms likely play a role. A number of internal galaxy transformation processes scale with stellar mass. Radiative and mechanical feedback from AGN can displace cold gas from the disc and heat central gas, suppressing gas cooling and subsequent star formation \citep[e.g.][]{Bower2006,Croton2006}. Recent semi-analytic models and hydrodynamic simulations \citep[e.g.][]{Springel2005,Hopkins2006,Dubois2014,Sijacki2015} argue that AGN feedback prescriptions are required to suppress star formation. Bar features in galaxies can efficiently drive gas into nuclear regions, inducing central star formation thereby reducing gas consumption time-scales \citep[e.g.][]{Knapen1995,Sheth2005, Cheung2013} and spurring (pseudo)bulge growth \citep{Kormendy2004,Athanassoula2005}. Morphological quenching, wherein the growth of a stellar bulge stabilizes the gas disc against fragmentation \citep{Martig2009} can further inhibit star formation. Finally, outflows driven by supernova feedback can eject cold gas from the galactic disc, suppressing star formation \citep[e.g.][]{Stinson2013,Hopkins2014,Keller2016}. 

The environmental mechanisms that are known to influence galaxy evolution are also numerous. A galaxy within the virial radius of a host halo can experience ram pressure stripping of cold gas from its disc \citep{Gunn1972}, reduced accretion of hot gas \citep[e.g.][]{Larson1980,Balogh2000,Kawata2007}, high-speed tidal interactions with nearby galaxies \citep{Moore1996, Moore1998} as well as mergers \citep{Toomre1972, Makino1997}, all of which can suppress star formation over varying time-scales. Although first discovered in galaxy clusters, subsequent works have found that the environmental trends at fixed stellar mass extend to groups \citep[e.g.][]{Postman1984, Zabludoff1998, Lewis2002, Gomez2003, McGee2011}. This suggests that not only does the dense core of a galaxy cluster affect galaxy properties, but satellite galaxies may undergo significant pre-processing in groups and may already have begun experiencing environmental effects well before entering the cluster environment \citep[e.g.][]{Kodama2001, Balogh2002, Hou2014, Roberts2017}. It is therefore important to understand the relative importance of these mechanisms not just in massive clusters but in low mass groups as well.

Although most galaxies are either blue, star-forming and disc-dominated or red, quiescent and early-type, there are exceptions. One way to investigate the mechanisms that dominate galaxy evolution is to \textcolor{black}{study} objects outside of these two main populations. These exceptions may be transitional objects evolving from active and blue to passive and red or they may be one stage of a more complicated evolutionary picture. Numerous studies have examined the significant population of so-called `passive red spirals', i.e. galaxies with late-type morphologies but red optical colours \citep[e.g.][]{Poggianti1999, Poggianti2004, Goto2003, Masters2010, Fraser2016}.  Other examples of outlier populations in the literature include blue, late-type passive galaxies \citep[e.g.][]{Mahajan2009}, blue, star-forming early-type galaxies \citep[e.g.][]{Schawinski2009a, Huertas2010, Ilbert2010}, \textcolor{black}{and `green valley' galaxies which have colours intermediate between the blue cloud and the red sequence \citep[e.g.][]{Martin2007, Wyder2007, Schawinski2014}}.

The aim of this work is to characterize galaxies that are red in optical colour but exhibiting significant star formation. Several authors have studied red, star-forming galaxies using a variety of colour and star formation metrics over a range of samples \citep[e.g.][]{Hammer1997, Coia2005, Demarco2005, Wolf2005, Wolf2009, Davoodi2006, Weinmann2006, Popesso2007, Koyama2008, Koyama2011,  Gallazzi2008, Saintonge2008, Haines2008, Verdugo2008, Brand2009, Mahajan2009}. However, the majority of the above works restricted their analysis to the cluster environment and few controlled for stellar mass to disentangle internal and environmental \textcolor{black}{effects}. These red star-forming galaxies have never been the sole focus of a study taking full advantage of the sample size and range of environments in the Sloan Digital Sky Survey and related data products.

In this work we investigate red, star-forming galaxies, hereafter referred to as Red Misfits, comparing them to the blue-and-active and red-and-quiescent populations in the Sloan Digital Sky Survey Data Release 7 \citep[SDSS DR7;][]{Abazajian2009}. In particular, we examine the stellar mass and morphology distributions of Red Misfits as well as their dust content, AGN abundance and preferred environments as defined by their host group halo mass and halo-centric radius. We determine that Red Misfits represent a physically-distinct population separate from the typical blue-and-active and red-and-quiescent populations and can help constrain quenching mechanisms and time-scales.

In Section 2 we describe our sample and the inclination-dependent colour corrections we apply. The properties of Red Misfits\textcolor{black}{, including stellar mass, morphology, dust content, gas-phase metallicity, AGN abundance and environmental trends} are described in Section 3. The discussion of our findings are presented in Section 4 and we summarize in Section 5. Throughout this work we assume a flat ${\Lambda}$CDM cosmology with ${\Omega_m}$=0.3, $\Omega_{\Lambda}$=0.7, $h$=0.7.   

\section{Sample} \label{Sample}
\subsection{Full Galaxy Catalogue} \label{TotalCat}

Our galaxy catalogue derives from the New York University Value-Added Galaxy Catalog \citep[NYUVAGC;][]{Blanton2005b} based on the Sloan Digital Sky Survey Data Release 7 \citep[SDSS DR7;][]{Abazajian2009}. Specifically we use the \textit{ugrizJHK} Petrosian absolute magnitudes, k-corrected to z=0.1 and corrected for Milky Way dust extinction \citep{Schlegel1998} based on the \texttt{kcorrect} code of \citet{Blanton2007a}. We use emission line measurements from the Max Planck Institut f{\"u}r Astrophysik and Johns Hopkins University (MPA-JHU) collaboration\footnote{http://www.mpa-garching.mpg.de/SDSS/DR7/}. \textcolor{black}{Based on these measurements, the MPA-JHU collaboration also provides specific star formation rates \citep[sSFRs,][]{Brinchmann2004} and stellar masses \citep{Kauffmann2003a} for the galaxies in our sample. Both the sSFR and stellar mass estimates include updated prescriptions from \citet{Salim2007} to correct for fiber aperture effects and contamination due to active galactic nuclei (AGN) using far- and near- UV photometry from the \textit{GALEX} mission \citep{Martin2005}}. In the interest of completeness we restrict our sample to galaxies with stellar masses above 10$^{9.5}$ M$_{\sun}$ and redshifts below 0.1.

\textcolor{black}{We obtain structural parameters for each galaxy by matching our sample to the catalogue of \citet{Simard2011} who use the \texttt{GIM2D} software \citep{Simard2002} to fit surface brightness profiles with parametric models}. Three fits are provided: an n$_b$=4 de Vaucouleurs bulge+disc decomposition, a free n$_b$ bulge+disc decomposition and a single-component pure S\'{e}rsic decomposition. We use the results from the single-component pure S\'{e}rsic decomposition in this study.

This sample is not volume-limited; as a result, it will suffer from the Malmquist bias \citep{Malmquist1925}, leading to a bias towards higher-luminosity objects that increases with redshift. To correct for this we weight by V$_{max}$, the comoving volume of the universe out to \textcolor{black}{the maximum} comoving radius at which the galaxy would have met the selection criteria of the sample. V$_{max}$ values are included in the \citet{Simard2011} catalogue. To prevent the contamination of our sample by extremely high-weight galaxies we remove from our sample the very \textcolor{black}{small number of} galaxies with V$_{max}$<1 Mpc\textcolor{black}{$^3$}.

\subsection{Colour Correction} \label{Correction}

We correct for inclination-induced colour bias in our sample following the general method of \citet{Maller2009}. In short, the magnitudes are corrected such that the average observed z=0.1 k-corrected $g-K$ and $r-K$ colours of inclined galaxies match the \textcolor{black}{average $g-K$ and $r-K$ colours of face-on galaxies at fixed M$_K$ and S\'{e}rsic index.} We make modifications to the \citet{Maller2009} methodology to achieve better results with our sample. Most importantly, we adopt a different function to fit the \textcolor{black}{$g-K$ and $r-K$ colours of face-on galaxies} to $M_K$ and S\'{e}rsic index. See Appendix \ref{App:Maller} for a more detailed explanation of our correction method. It is important to note that this correction is not intended to recover the dust-unaffected colours of galaxies, rather it recovers the face-on colours of galaxies. The method simply reverses the additional effects of inclination on colour. The inclination-corrected colours of a sample of galaxies may still be significantly dust-reddened if there is a large amount of intrinsic dust within the sample (see Section \ref{Dust}).

To allow for accurate colour corrections we remove the $\sim$0.1 per cent galaxies in our sample with missing K-band magnitudes or S\'{e}rsic indices. Our full sample therefore consists of 277785 galaxies after all cuts have been applied. Inclination and z=0.1 k-corrected $g-r$ colours are denoted (M$_g^i$ - M$_r^i$)$^{0.1}$ for the remainder of this work.

\subsection{Emission Line Subsample} \label{emission}

To determine the amount of intrinsic dust reddening (see Section \ref{Dust}) and the abundance of AGN (see Section \ref{AGN}) in star-forming, red galaxies as compared to the \textcolor{black}{larger blue-and-star-forming and red-and-quiescent populations}, we require a sample of galaxies with reliable emission line strength measurements. We construct an emission-line subsample by selecting the galaxies in our full sample with  \textcolor{black}{signal-to-noise of 3} or greater in H$\alpha$, H$\beta$, N\textsc{ii} $\lambda_{6584}$ and O\textsc{iii} $\lambda_{5007}$. We also require z>0.05 to reduce the aperture bias of the SDSS fibre. \textcolor{black}{The 3" fibre corresponds to $\sim$3 kpc at z=0.05, compared to a typical half-light radius of $\sim$4 kpc along the semi-major axis for galaxies at the same redshift.} This emission line subsample is $\sim$1/3 of the total sample and is heavily biased towards galaxies with significant star formation and/or AGN components (see Section \ref{Classification}). 

\subsection{Environmental Catalogues}

\subsubsection{Group Catalogue} \label{group}

To study the distribution of Red Misfits in different environments in the Universe, we match our full catalogue to the DR7 group catalogue of \citet[][hereafter Y07]{Yang2007}. The iterative algorithm for halo-based group finding is outlined in detail in \citet{Yang2005a} and Y07. The stellar mass of each group is defined as 
\begin{center}
	\begin{equation}
		\scalebox{1.4}{$M_{*,grp} = \frac{1}{g(L_{19.5},L_{lim})} \sum_{i}\frac{M_{*,i}}{C_i}$ \text{,}}
	\end{equation}
\end{center}
where M$_{*,i}$ is the stellar mass of the i'th galaxy in the group, C$_i$ is the completeness of the survey at the position of that galaxy, and g(L$_{19.5}$,L$_{lim}$) is a correction factor accounting for the magnitude limit of the survey. The mass of each group's halo is estimated using abundance matching to M$_{*,grp}$ with the halo mass function of \citet{Warren2006} and the transfer function of \citet{Eisenstein1998}.

After removing the 65.6 per cent of galaxies residing in \textcolor{black}{single systems or groups too small to have halo masses assigned,} our final group sample consists of 95648 galaxies in 29346 groups with halo masses ranging from $10^{11.75}$M$_{\sun}$ to $10^{15.1}$M$_{\sun}$. The highest mass systems are galaxy clusters, but we will refer to all galaxies in this catalogue as group galaxies for simplicity.

We use projected group-centric distance and halo mass as measures of galaxy environment, both of which have been shown to correlate with local overdensity \citep{Peng2012,Haas2012,Woo2013,Carollo2013}. We calculate the projected group-centric distance \textcolor{black}{(defining the group centre as the location of the most massive galaxy in the group)} using the methods outlined in \citet{Hogg1999} and normalize to r$_{200}$ given by \citet{Tinker2008} and Y07:

\begin{center}
  \begin{equation}
  	 \scalebox{1.3}{$r_{200}=\left(\frac{3M_{halo}}{4 \pi (200 \overline{\rho_m})} \right)^{1/3}$ ,}
  \end{equation}
\end{center}
which in our cosmology becomes

\begin{center}
	\begin{equation}
		\scalebox{1.3}{$r_{200}=1.13 h^{-1} \left(\frac{M_{halo}}{10^{14}h^{-1}M_{\odot}} \right)^{1/3}(1+z_{group})^{-1}$ ,}
	\end{equation}
\end{center}
where z$_{group}$ is the redshift of the centre of the group.

\begin{figure*}
  \includegraphics[width=0.48\textwidth]{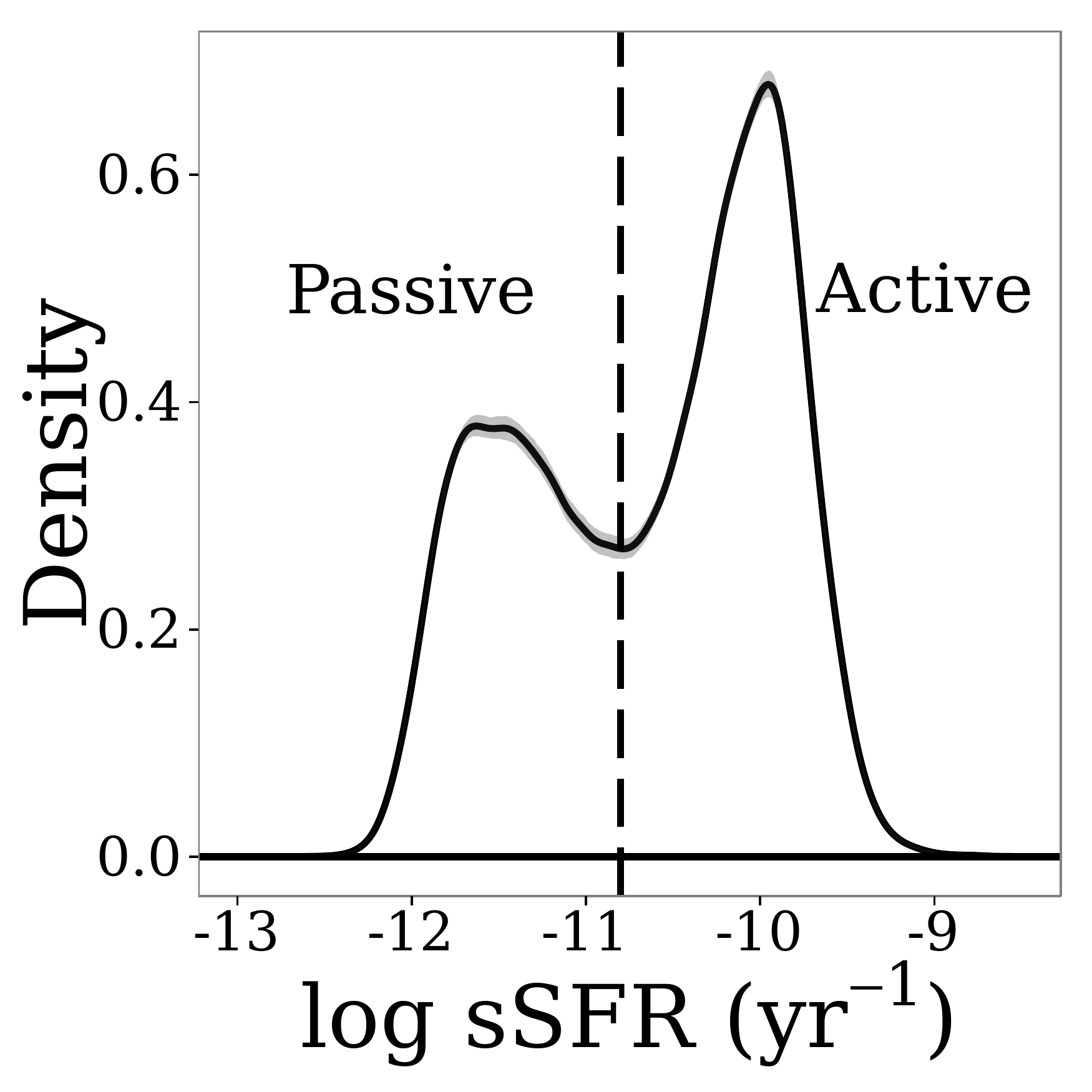} 
  \includegraphics[width=0.48\textwidth]{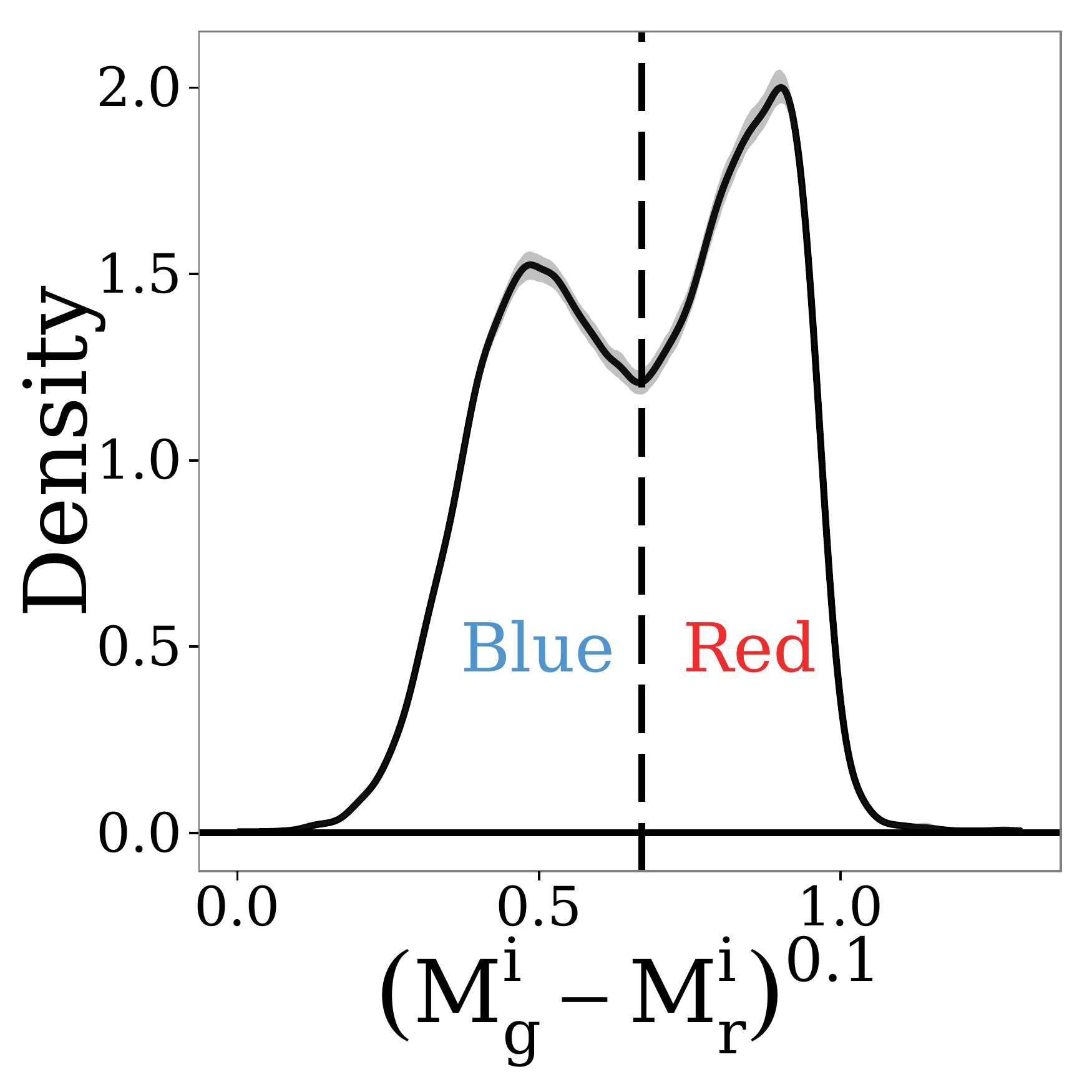}
  \caption{\textit{Left}: V$_{max}$-weighted specific star formation rate (sSFR) Gaussian kernal-smoothed distribution for galaxies in the full sample. The local minimum at log(sSFR)=-10.8 yr$^{-1}$ defines our `active' and `passive' samples. \textit{Right}: V$_{max}$-weighted (M$_g^i$ - M$_r^i$)$^{0.1}$ colour distribution of galaxies in the full sample. A cut at (M$_g^i$ - M$_r^i$)$^{0.1}$=0.67 mags defines our `red' and `blue' populations. Shaded regions in both plots show 99\% confidence intervals from 1000 bootstrap resamplings.}
  \label{fig:histograms}
\end{figure*}

\subsubsection{\textcolor{black}{Isolated Galaxy Catalogue}} \label{field}

As a companion to the group catalogue, we identify a catalogue of isolated field galaxies to probe the population of Red Misfits in low-density environments. We construct our isolated field catalogue using the N=1 groups in the Y07 catalogue with an additional isolation criterion: a galaxy must have no neighbour within 1 Mpc and 1000 kms$^{-1}$ brighter than the SDSS magnitude limit at our redshift cut of z=0.1. To eliminate the possibility that galaxies in this catalogue have close neighbours just outside our sample region, we remove galaxies within 1 Mpc of the SDSS coverage boundary or within 1000 km s$^{-1}$ of z=0.1 as in \citet{Roberts2017}. Our isolated sample consists of 112614 galaxies.

\subsection{Galaxy Classification} \label{Classification}

\begin{figure*}
  \includegraphics[width=\textwidth]{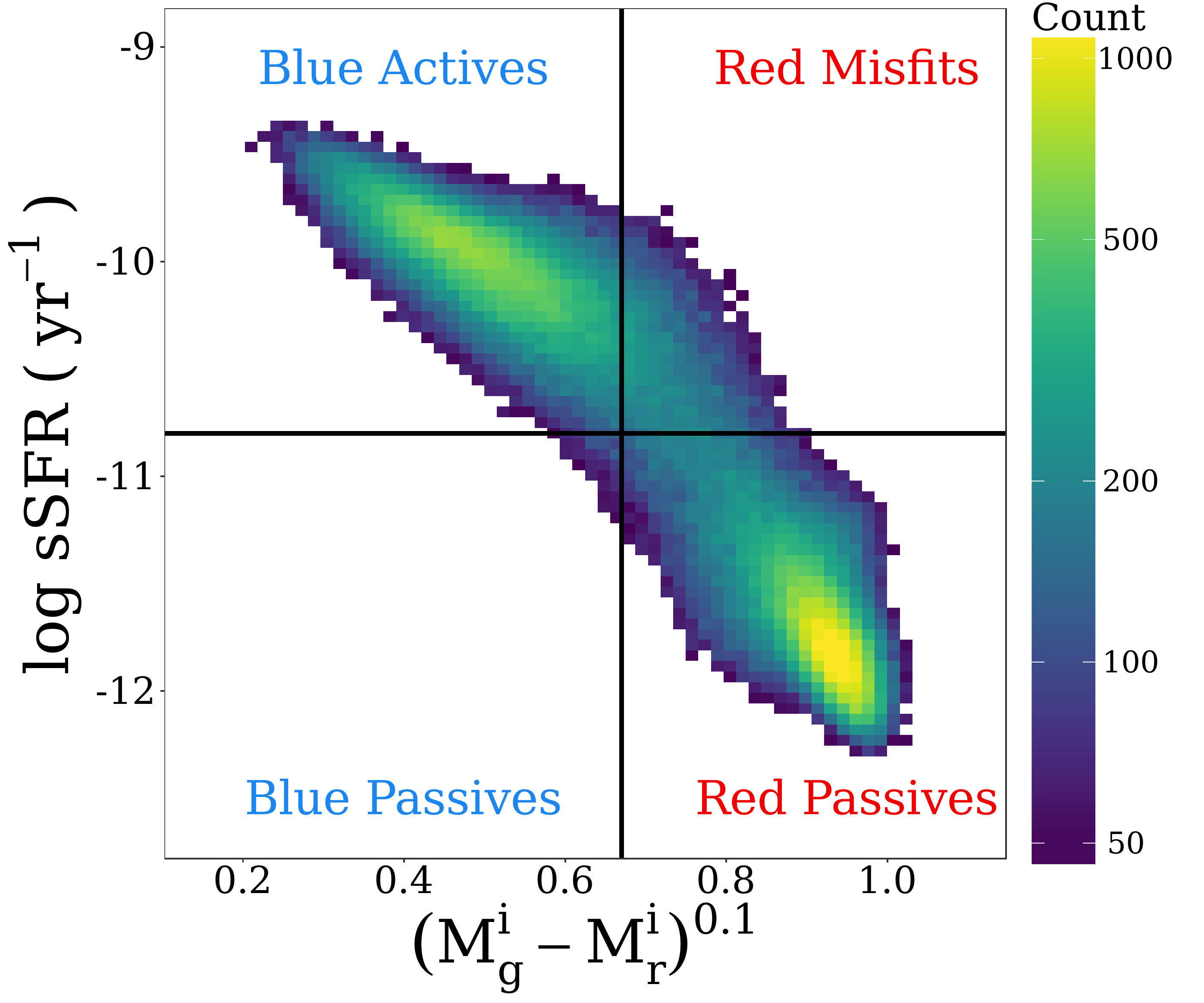}
  \caption{sSFR vs. inclination-corrected and k-corrected colour for galaxies in the full sample. Red/blue and active/passive cuts are shown. The Blue Active, Blue Passive, Red Active (Red Misfit) and Red Passive quadrants are shown based on divisions shown in Fig. \ref{fig:histograms}. Bins populated by fewer than 50 galaxies are not shown.} 
  \label{fig:ssfrgr}
\end{figure*}

\begin{table*}

    \begin{tabular}{ c  c  c  c  c }
     & \multicolumn{4}{  c  }{Percent of Sample} \\ \cline{2-5}
     Population &  Full Sample & Emission-line Sample & Group Sample & Isolated Sample \\
      & (277785 galaxies) & (90000 galaxies) & (95648 galaxies) & (112614 galaxies) \\ \hline 
      Blue Active & 42.3\% & 74.7\% & 27.2\% & 52.2\% \\ \hline
      Blue Passive & 1.7\% & 1.0\% & 1.6\% & 1.7\% \\ \hline
      Red Active & 10.9\% & 15.4\% & 11.0\% & 10.9\% \\
      (Red Misfit) & & & & \\ \hline
      Red Passive & 45.1\% & 8.8\% & 60.2\% & 35.2\% \\ \hline
    \end{tabular} 
  \caption{Populations of each of the four galaxy populations in the four samples. }
  \label{Table 1}
\end{table*}

We separate galaxies into four populations based on their inclination-corrected colour and specific star formation rate. V$_{max}$-weighted sSFR and colour density \textcolor{black}{distributions} can be seen in Fig. \ref{fig:histograms}. The sSFR distribution shows clear bimodality with peaks near $10^{-11.5}$ yr$^{-1}$ and $10^{-10}$ yr$^{-1}$ and a local minimum at $\approx$10$^{-10.8}$ yr$^{-1}$. We use this minimum to divide our sample into `active' galaxies (sSFR>10$^{-10.8}$ yr$^{-1}$) and `passive' ones (sSFR<10$^{-10.8}$ yr$^{-1}$). \textcolor{black}{This bimodality in sSFR and the break in the vicinity of sSFR$\approx$10$^{-11}$ yr$^{-1}$ has been noted by several authors \citep[e.g.][]{Brinchmann2004,Kauffmann2004,Wetzel2012} and is largely independent of stellar mass and environment \citep{Wetzel2012}. Defining active and passive subsamples based on a fit to the star-forming main sequence \citep[see e.g.][]{Brinchmann2004, Noeske2007b, Elbaz2007} would yield similar results -- only 3 per cent of our `active' galaxies are more than 2$\sigma$ below the main sequence of star formation.}

\textcolor{black}{\citet{Brinchmann2004} derive specific star formation rates by constructing a grid of $\sim$2x10$^5$ stellar synthesis models \citep{Bruzual1993, Bruzual2003} spanning a range in metallicity, dust-to-metal ratio, dust attenuation and ionization parameter. The model emission line ratios are compared to the observed spectrum and log likelihood functions of each model for each galaxy are constructed and used to determine the probability distribution function (PDF) for the sSFR of each galaxy. In this work we use the median of each PDF as our estimate for sSFR. The median uncertainty of each sSFR estimate for our star-forming (sSFR>10$^{-10.8}$ yr$^{-1}$) galaxies is $\sim$0.3 dex. Galaxies with undetected emission lines are compatible with a wide range of low star formation rates, therefore the sSFRs of the quiescent galaxies in the vicinity of $10^{-12}$ yr$^{-1}$ should be viewed as an upper limit.} While the active peak of the sSFR distribution in Fig. \ref{fig:histograms} is physical, the height of the quiescent peak is artificial, as the true distribution should tail off to lower sSFRs \citep{Salim2007}. 

The distribution of inclination-corrected Petrosian (M$_g^i$ - M$_r^i$)$^{0.1}$ colours can also be seen in Fig. \ref{fig:histograms}. The distribution is bimodal, in agreement with previous studies \citep[e.g.][]{Strateva2001,Baldry2004,Baldry2006,Balogh2004b,Blanton2005a,Coil2008}. We divide our sample into `red' and `blue' populations \textcolor{black}{using the local minimum at} (M$_g^i$-M$_r^i$)$^{0.1}\approx$ 0.67 mags. Defining our red and blue populations by fitting a red sequence \textcolor{black}{in inclination-corrected colour-magnitude space} does not change any of our results, as the slope of the red sequence is negligible.

Fig. \ref{fig:ssfrgr} shows the distribution of our sample in sSFR - intrinsic colour space with our sSFR and inclination-corrected colour cuts overlaid. We classify galaxies by the quadrant they occupy in Fig. \ref{fig:ssfrgr}: Blue Actives (blue and star-forming), Red Misfits (red and star-forming), Red Passives (red and quiescent) and Blue Passives (blue and quiescent). The \textcolor{black}{relative sizes of each population in each of our samples are listed in Table \ref{Table 1}}. $\sim$87 per cent of galaxies in our full sample belong to either the Blue Active or Red Passive populations while a significant minority ($\sim$11) per cent of galaxies are Red Misfits. \textcolor{black}{For the remainder of this work we do not consider the small population of galaxies ($\sim$2 per cent of the full sample) classified as Blue Passives; we instead focus on the properties of Red Misfits compared to Blue Actives and Red Passives and defer Blue Passives to future work.}

We note that while these subpopulation sample sizes change depending on the colour and sSFR cuts used, the results in Section \ref{Results} are robust against the precise locations of these cuts. 

\section{Results} \label{Results}
\subsection{Stellar Mass}

\begin{figure}
 \includegraphics[width=0.5\textwidth]{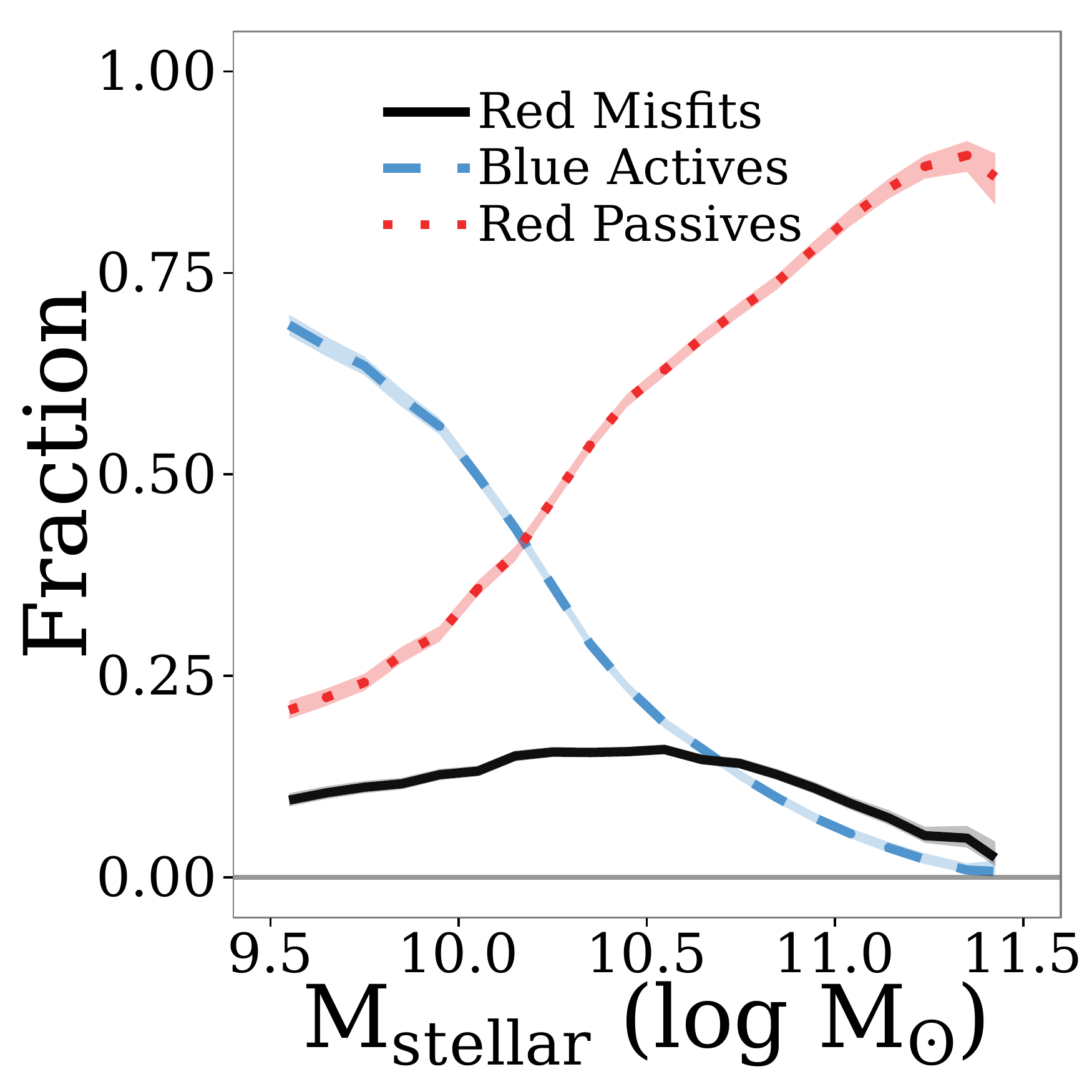}
  \caption{V$_{max}$-weighted relative populations of Blue Active, Red Passive and Red Misfit galaxies in the full sample binned by stellar mass. Shaded regions show 99\% confidence intervals generated by the beta distribution as outlined by \citet{Cameron2011}.} 
  \label{fig:mass}
\end{figure}

Many galaxy properties such as star formation rate \citep[e.g.][]{Salim2007,Noeske2007a,Poggianti2008}, colour \citep[e.g.][]{Baldry2006,vandenBosch2008a,Bamford2009} and morphology \citep[e.g.][]{vanderWel2008,Bamford2009,Bluck2014} correlate with stellar mass; \textcolor{black}{throughout this work} we therefore compare the properties of Red Misfits, Blue Actives and Red Passives at fixed stellar mass. Fig. \ref{fig:mass} shows the relative populations of Blue Active, Red Passive and Red Misfit galaxies in different bins of stellar mass in the full sample. \textcolor{black}{As expected, Blue Active galaxies dominate at low stellar mass while at high stellar mass Red Passives are the most common galaxy population. The proportion of Red Misfits, however, depends only weakly on stellar mass. They comprise $\sim$10 per cent of our sample at low stellar mass, peaking at $\sim$16 per cent at M$_{stellar}$ $\simeq$ $10^{10.5}$ $M_{\sun}$ before falling to $\sim$5 per cent at high stellar mass. }

\subsection{Structure \& Morphology }

\begin{figure*}
	\centering
  \includegraphics[width=0.49\textwidth]{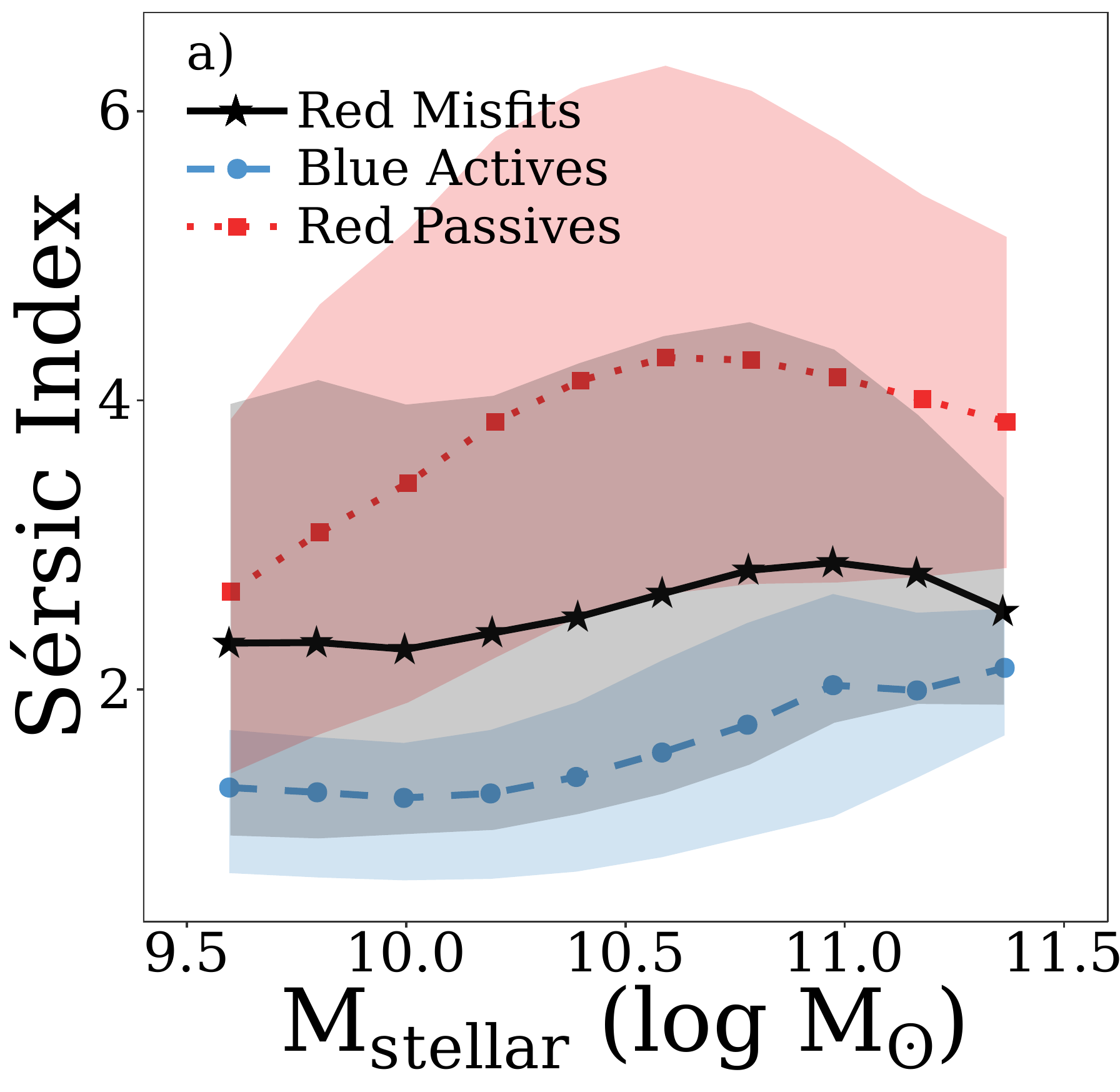}
\includegraphics[width=0.49\textwidth]{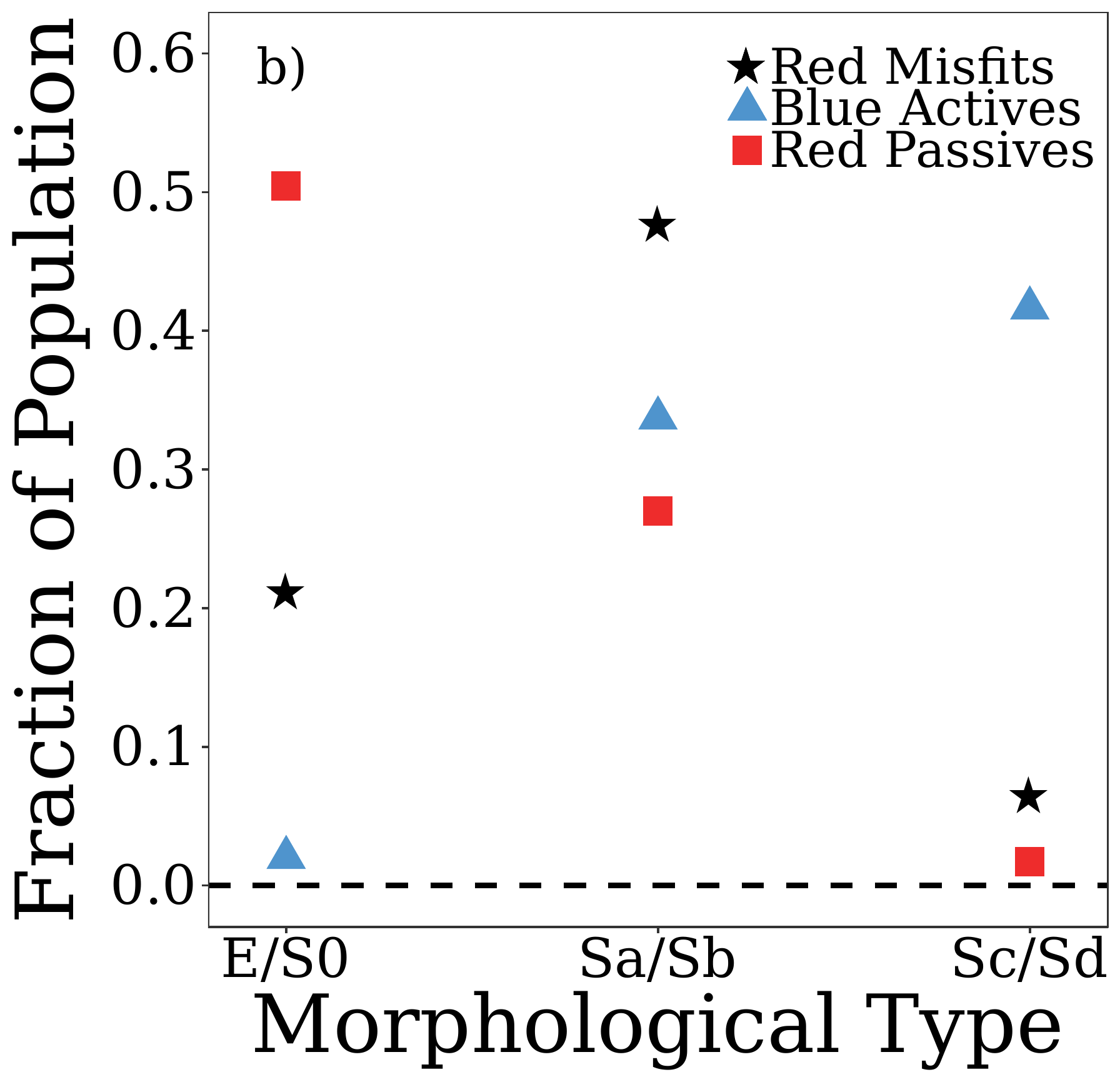}
  \caption{\textit{a)}: Mean S\'{e}rsic index of Blue Active, Red Passive and Red Misfit galaxies against stellar mass.  \textcolor{black}{Symbols indicate V$_{max}$-weighted mean value}. Shaded regions span the \textcolor{black}{unweighted} 16th to 84th percentiles of each bin. \textcolor{black}{\textit{b)}: Distribution of Blue Active, Red Passive and Red Misfit morphological types \citep{Huertas2011} in our full sample. Red Misfits show intermediate, early-spiral morphologies. Error bars are smaller than symbols in all cases.} }
  \label{fig:morphhist}
\end{figure*} 

The correspondence between colour and morphology \citep[e.g.][]{deVaucouleurs1961, Strateva2001, Baldry2004} as well as star formation rate and morphology \citep[e.g.][]{Kauffmann2003b, Bell2012, Woo2015} indicate that the same processes that control a galaxy's star formation history may also control its structure. In Fig. \ref{fig:morphhist}a we show the mean S\'{e}rsic index of the Red Misfit, Blue Active and Red Passive populations in the full sample binned by stellar mass. When using the S\'{e}rsic index as a structural proxy for morphology, the Red Passive population has more elliptical (higher-index) morphologies than the Blue Active population. Red Misfits are intermediate in structure between the Blue Active and Red Passive populations at all stellar masses and their morphologies exhibit the weakest dependence on stellar mass out of the three populations. Qualitative results remain the same if we use alternate structural proxies for morphology such as concentration (c$\equiv$r$_{90}$/r$_{50}$) or bulge-to-total light ratio: Red Misfits are more (less) bulge-dominated and concentrated than Blue Actives (Red Passives).

Structural proxies for galaxy morphology such as the S\'{e}rsic index provide a quantified metric of what is a complex galaxy property. The price of using these proxies is that the more subtle morphological features of each individual galaxy are lost and each proxy can introduce its own bias to the sample. For example, bulge light can dominate the surface brightness profiles of late-type galaxies with prominent bulges, therefore they may have structural morphology proxies more typical of early-type galaxies \citep[e.g.][]{Hameed2005,Cheng2011}. When available, morphological classification for each galaxy can provide extra insight over using these structural proxies.

In addition to the S\'{e}rsic index, we probe morphology using the classifications of \citet{Huertas2011}, hereafter HC11, obtained using an automated Bayesian approach with the classifications of \citet{Fukugita2007} as a training sample. Each galaxy is assigned a probability, $p_{class}$ of being in one of four morphological classes (E, S0, Sab, Scd). While more detailed visual morphological classifications are available \citep[e.g.][]{Nair2010, Baillard2011, Willett2013}, these catalogues classify a small and biased subset of our full sample. In contrast, 275173 galaxies in our full sample (99\%) have HC11 morphological classifications.

Fig. \ref{fig:morphhist}b shows the distribution of HC11 morphological classifications for Red Misfits, Blue Actives and Red Passives. We adopt a p$_{class}$>0.5 threshold to decide whether a galaxy belongs to a given class. 78\% of our sample is classified as one of the four types under this threshold while the remaining 22\% do not have p$_{class}$>0.5 in any single class. The HC11 classifications mirror the trends seen in the S\'{e}rsic indices -- Red Passives are most often classified as ellipticals or S0's by HC11 and Blue Actives as late spirals. Red Misfits, however, are most likely to be classified as early spirals. Therefore in both visual and structural morphology Red Misfits have predominantly intermediate morphologies.

\subsection{Dust Content} \label{Dust}

\begin{figure}
  	  \includegraphics[width=0.5\textwidth]{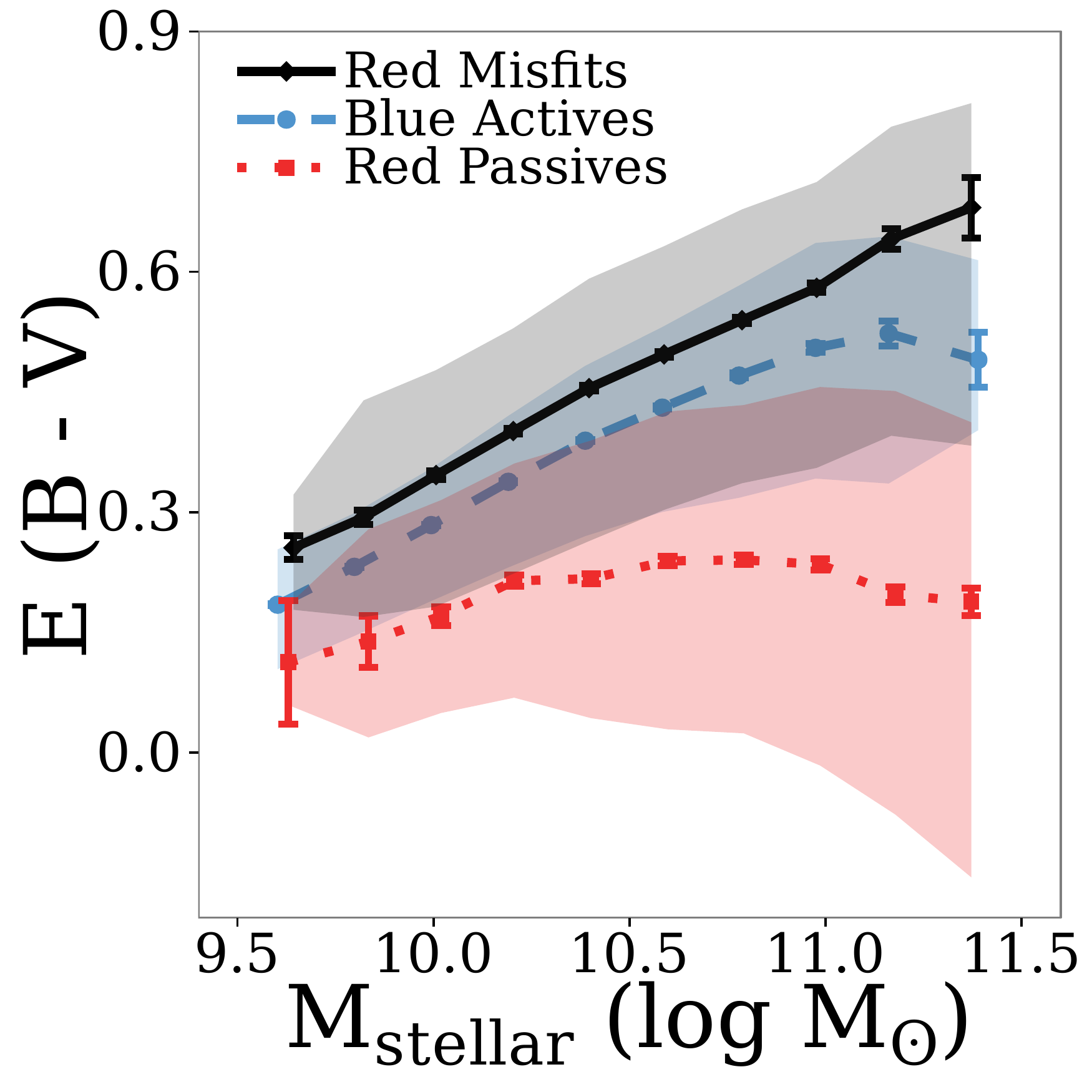}
      \caption{Optical colour excess due to intrinsic dust against stellar mass for non-edge-on (e<0.75) Blue Active, Red Passive and Red Misfit galaxies in the emission-line sample. Error bars indicate 1$\sigma$ errors on the V$_{max}$-weighted mean value of each bin. Shaded regions span the 16th to 84th percentile of each bin.} 
  \label{fig:Balmer}
\end{figure}

We have selected Red Misfits using inclination-corrected colours to remove the effect of inclination on optical colours, but the possibility still exists that Red Misfits have been significantly reddened due to large amounts of intrinsic dust. Dusty, star-forming red galaxies have been widely studied in the literature \citep[e.g.][]{Coia2005, Wolf2005, Wolf2009, Gallazzi2008, Saintonge2008, Brand2009}. In this section we explore the fraction of Red Misfits whose colours could be explained with significant reddening by intrinsic dust.

We quantify the amount of dust extinction using the Balmer decrement, \textcolor{black}{the ratio of H$\alpha$ flux to H$\beta$.} It is possible to determine the expected ratio between these lines \textcolor{black}{based} on the properties of the absorbing medium \citep{Menzel1937,Baker1938}. The flux ratio depends weakly on gas temperature and density \citep{Osterbrock2006} but is strongly affected by dust. \textcolor{black}{The ionized gas colour excess in $B-V$ due to dust is characterized by}

\begin{center}
\begin{equation}
	\scalebox{1.3}{$E(B-V) = 1.97\log\left(\frac{(H\alpha / H\beta)_{obs}}{2.86}\right)$ ,}
\end{equation}
\end{center}
where $(H \alpha/H \beta)_{obs}$ is the observed \textcolor{black}{Balmer decrement, 1.97 is a constant derived using the reddening curve of \citet{Calzetti2000} and 2.86 is the unreddened flux ratio \citep{Osterbrock1989}.} The choice of 2.86 for the unreddened flux ratio is common in the literature \citep[e.g.][]{Xiao2012,Dominguez2013}. We note that since Balmer decrements are measured based on the H$\alpha$ and H$\beta$ emission lines, extinction is estimated based solely on the \textcolor{black}{emission-line} regions of a galaxy and therefore may not be representative of the galaxy as a whole.

In Fig. \ref{fig:Balmer} we plot $E(B-V)$ as a function of stellar mass for Blue Actives, Red Passives and Red Misfits. \textcolor{black}{To ensure reliable H$\alpha$ and H$\beta$ measurements we restrict our $E(B-V)$ analysis to our sample of emission-line galaxies (see Sec. \ref{emission})}. Since the Balmer decrements in our emission-line sample show a mild trend with axis ratio \citep[see also][]{Yip2010,Xiao2012}, we \textcolor{black}{omit from Fig. \ref{fig:Balmer} the 7.7 per cent of galaxies in our emission line sample that are edge-on (e>0.75). However, results are unchanged if the entire emission-line sample is used. The ionized gas} optical colour excess due to dust reddening is on average larger in Red Misfits than Blue Actives across the entire stellar mass range of our sample in Fig. \ref{fig:Balmer}. This enhancement in colour excess is small: Red Misfits \textcolor{black}{experience 0.066 more mags of reddening in $B-V$ for the ionized gas than Blue Actives on average across all stellar mass bins. Assuming the ratio of reddening in the stellar continuum to the reddening in the ionized gas is 0.44 \citep{Calzetti1997, Calzetti2000}, this corresponds to an extra stellar reddening in Red Misfits of 0.029 mags over Blue Actives in $g-r$ colour. Removing all Red Misfits within 0.029 mags of the colour cut in Fig. \ref{fig:ssfrgr} reduces our Red Misfit sample by $\sim$23.5 per cent but does not qualitatively change any results presented in this work.}

The scatter for all populations in Fig. \ref{fig:Balmer} is significant: only 26 per cent of Red Misfits exhibit colour excesses 1$\sigma$ over the colour excess for a typical Blue Active galaxy with the same stellar mass. The removal of these most-reddened Red Misfits does not qualitatively change any results presented in this work. We therefore conclude that although dust reddening can account for a small proportion of Red Misfits, the entire population cannot be dismissed as dust-reddened Blue Active galaxies. Even when the most conservative colour corrections are applied, a population of red star-forming galaxies persists.

\subsection{Stellar Age} \label{Age}

\begin{figure*}
	\centering
	\includegraphics[width=0.49\textwidth]{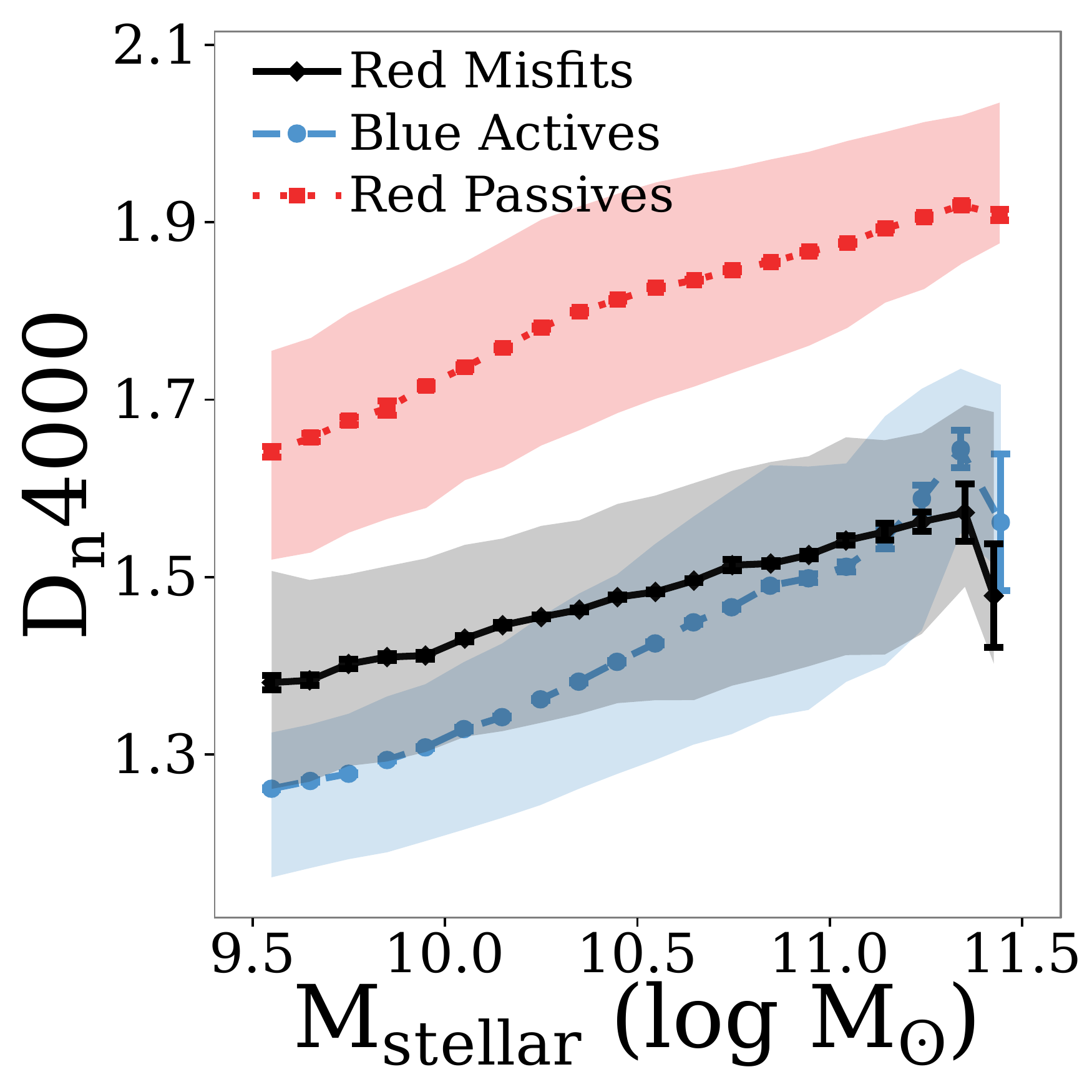}
	\includegraphics[width=0.49\textwidth]{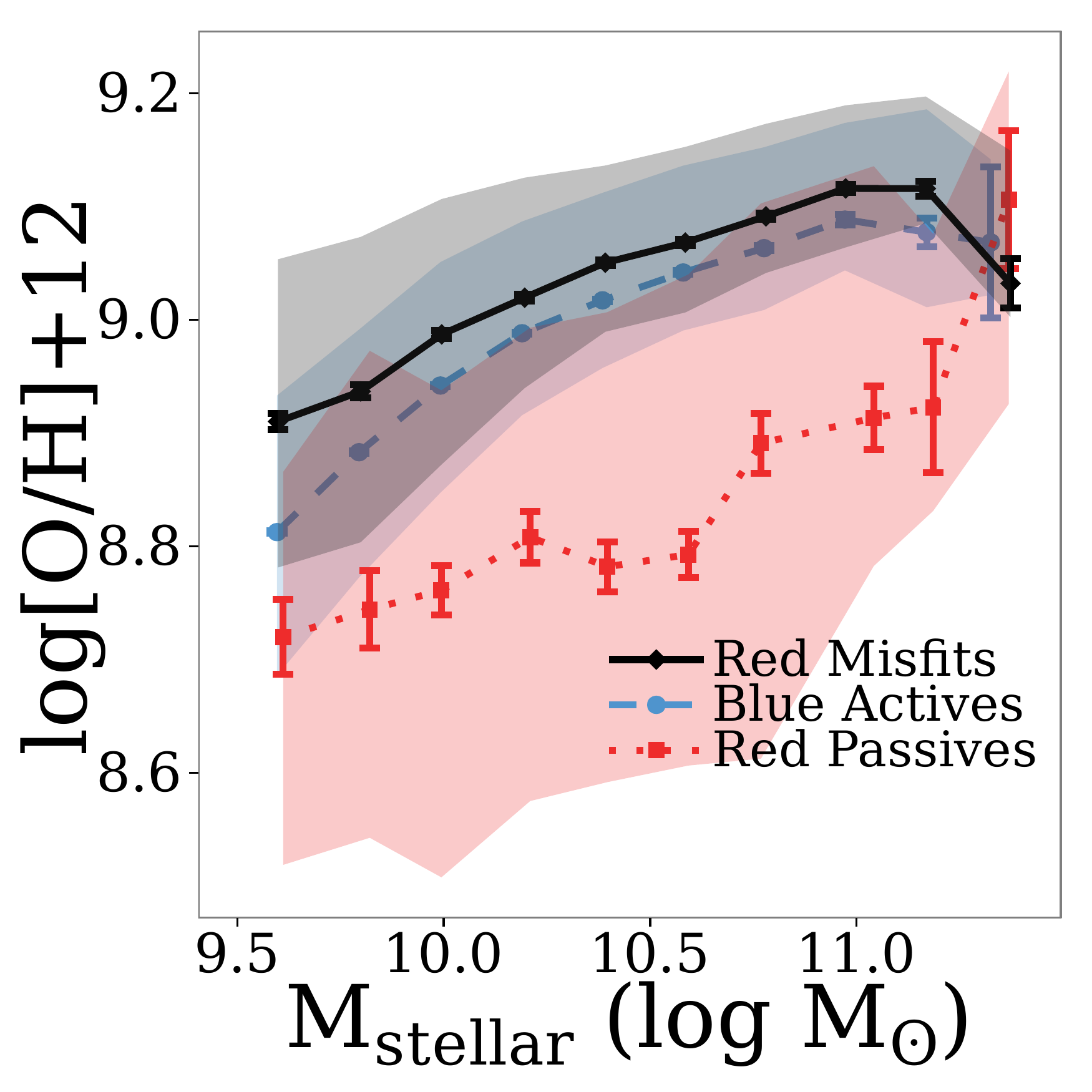}	
	\caption{\textit{Left:} D$_n$4000 index measurement for each population in the full sample binned by stellar mass. \textit{Right}: Gas-phase metallicity of each population in the full sample binned by stellar mass. Error bars in both plots indicate 1$\sigma$ errors on the V$_{max}$-weighted mean value. Shaded regions span the 16th to 84th percentile of each bin.}
	\label{fig:Dn4000}
\end{figure*}

Although Red Misfits are star-forming by construction, a stellar population that is overall older than the Blue Actives can significantly contribute to their red colours. To characterize the mean stellar age of each population we use the D$_n$4000 index as defined by \citet{Balogh1999} provided in the MPA-JHU emission-line catalogue. 

Fig. \ref{fig:Dn4000} shows the relationship between D$_n$4000 and stellar mass for Blue Actives, Red Passives and Red Misfits in the full sample. Red Passive galaxies show high values of D$_n$4000, confirming them as galaxies with an old stellar population and relatively little recent star formation. Conversely, Blue Actives exhibit low D$_n$4000 values, indicating that they have comparatively younger stellar populations. Both populations tend towards larger D$_n$4000 at higher stellar mass. At low and moderate stellar mass, Red Misfits exhibit an intermediate D$_n$4000 measurement between the Blue Active and Red Passive populations. Toward high stellar mass, the Red Misfit D$_n$4000 measurements approach the same value as the Blue Actives', indicating that their stellar populations are of comparable age. \textcolor{black}{There are several factors contributing to the red optical colours of Red Misfits -- their moderate sSFRs (see Fig. \ref{fig:ssfrgr}), their enhanced dust extinction and their aged stellar populations when compared to Blue Actives.}

\subsection{Gas-Phase Metallicity}

To probe metallicity we use the gas-phase metallicity (log[O/H]+12) measurements in the MPA-JHU catalogue based on \citet{Tremonti2004}. We stress that only a small fraction (30 per cent) of our full sample has metallicity measurements, as \citet{Tremonti2004} \textcolor{black}{measure gas-phase metallicity only for galaxies} with a 5$\sigma$ detection or greater in H$\alpha$, H$\beta$ and N\textsc{ii} $\lambda_{6584}$. The sample of galaxies with metallicity measurements \textcolor{black}{is therefore dominated} by Blue Actives ($\sim$90 per cent) and Red Passives are extremely underrepresented ($\sim$0.5 per cent). Fig. \ref{fig:Dn4000} shows the relationship between log[O/H]+12 for the Red Misfits, Blue Actives and Red Passives that meet these criteria. For all three populations, gas-phase metallicity increases with stellar mass. As the most metal-rich population by a slight margin, Red Misfits are on average $\simeq$0.03 dex more metal-rich than Blue Actives \textcolor{black}{at fixed stellar mass. This is consistent with recent studies finding an anti-correlation between sSFR and gas-phase metallicity at fixed M$_{stellar}$ for star-forming galaxies \citep[e.g.][]{Mannucci2010, Yates2012, Salim2014, Telford2016, Brown2016}}.

\subsection{AGN Abundance} \label{AGN}

\begin{figure*}
	  \includegraphics[width=\textwidth]{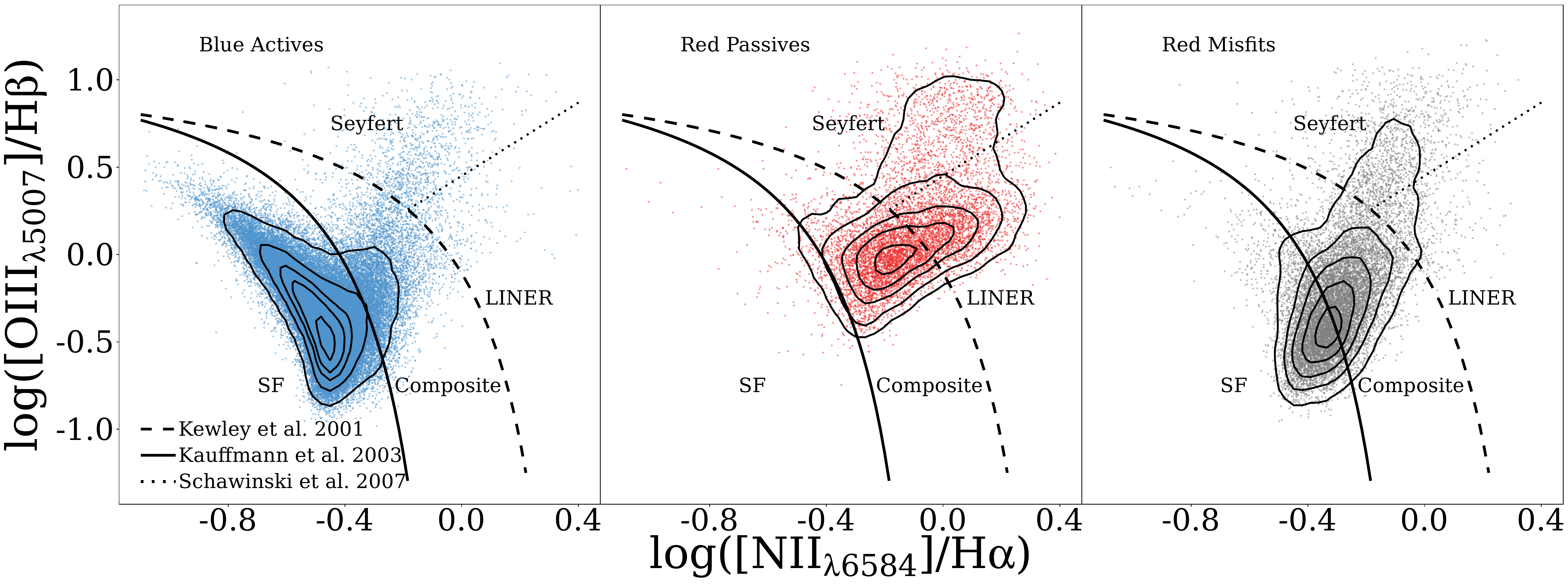}	  
      \caption{Distribution of Blue Actives (left), Red Passives (middle) and Red Misfits (right) in the emission-line sample on the BPT diagram. Contours encompass 10\%, 30\%, 50\%, 70\% and 90\% of the unweighted distributions. Lines from \citet{Kewley2001} and \citet{Kauffmann2003b} define star-forming and AGN regions of the diagram as well as the composite region where emission from stellar and non-stellar processes are comparable. \textcolor{black}{The dotted line from \citet{Schawinski2007} separates the AGN region into a Seyfert region and a LINER region.}} 
  \label{fig:BPT}
\end{figure*}

\begin{figure*}
	\centering
	  \includegraphics[width=0.49\textwidth]{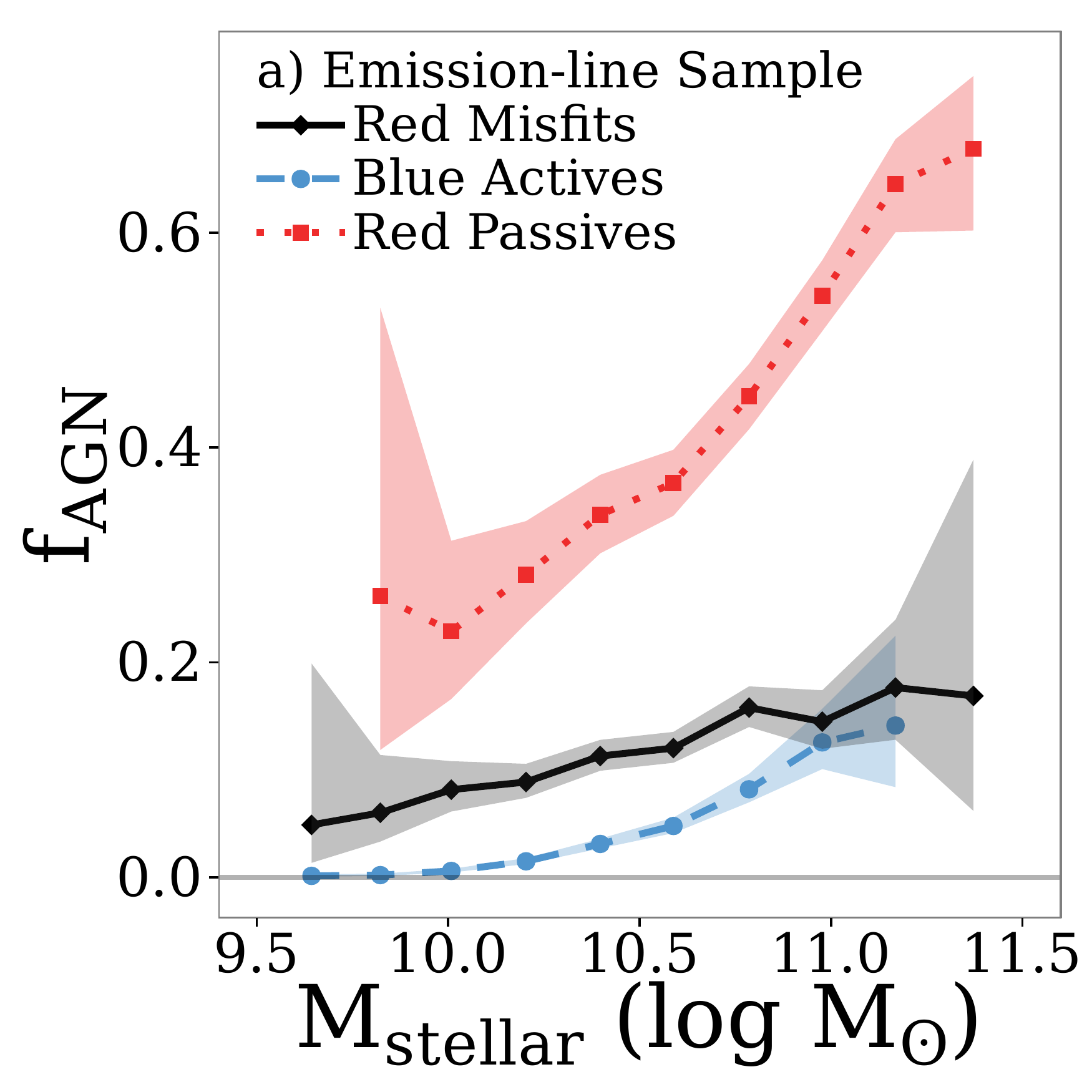}	
	  \includegraphics[width=0.49\textwidth]{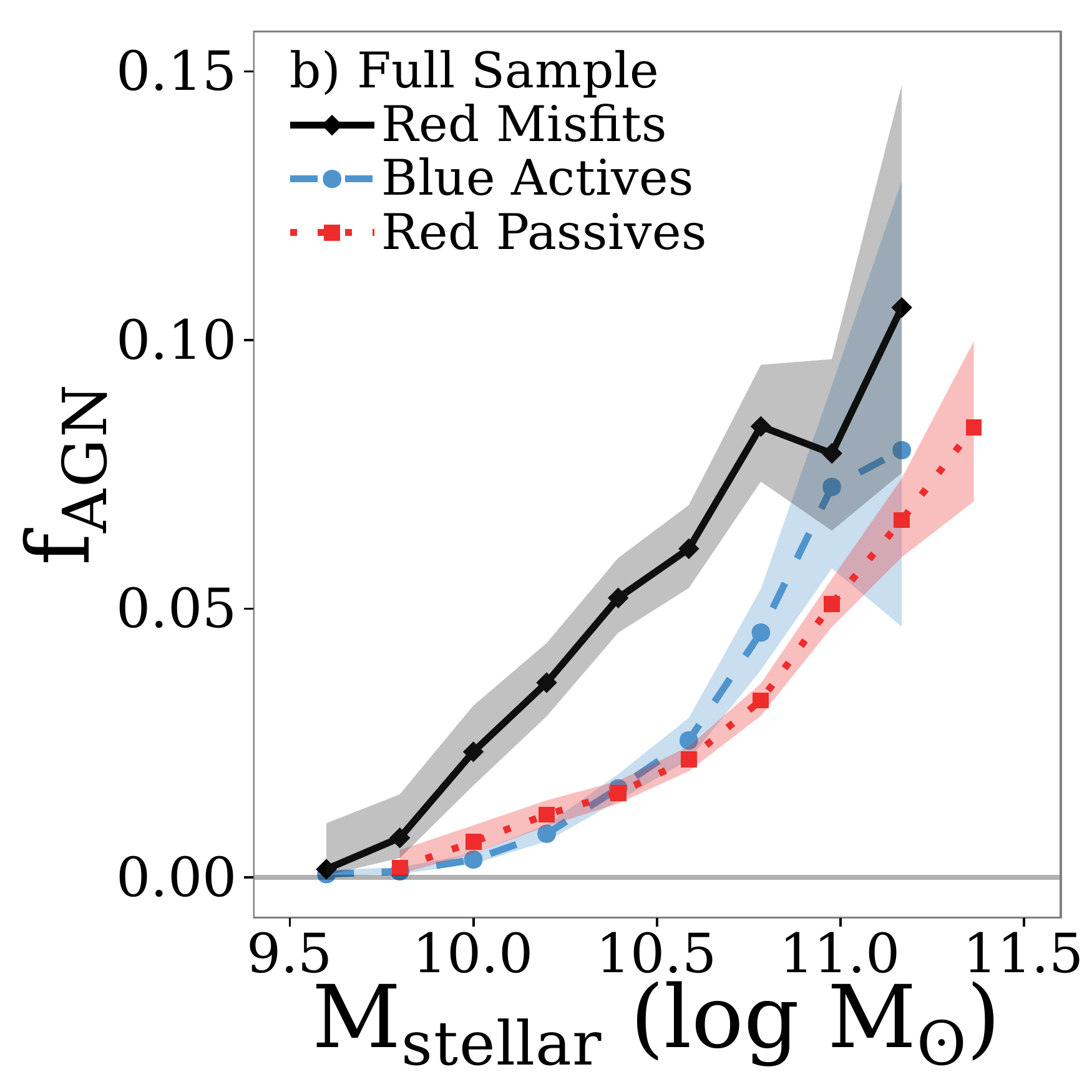}
      \caption{a) Fraction of each population in our emission-line subsample identified as hosting an AGN (i.e. fraction lying above the Ke01 line) in bins of stellar mass. b) AGN fraction of each population in our \textit{full} sample (i.e. \textcolor{black}{fraction of galaxies in the total sample that are both emission line galaxies and lie above the Ke01 line}). Shaded regions are 99\% confidence intervals generated by the beta distribution as outlined by \citet{Cameron2011}. }
  \label{fig:agn}
\end{figure*}

It is widely accepted that AGN can play a significant role in galaxy evolution, particularly for massive galaxies. To match observations in the local Universe, galaxy evolution models typically invoke AGN feedback -- heating central gas and suppressing star formation in massive galaxies \citep[e.g.][]{Granato2004,Bower2006}. Nuclear star formation associated with an AGN \citep[e.g.][]{Goto2006,Bildfell2008,Reichard2009} may help reconcile the significant star formation in Red Misfits with their red colours. 

Though AGN feedback can quench star formation in the long term, the link between \textit{ongoing} AGN emission and star formation is still debated, with various studies finding that AGN hosts experience enhanced star formation \citep[e.g.][]{Silverman2009, Rosario2015}, suppressed star formation \citep[e.g.][]{Salim2007,Gurkan2015,Ellison2016} or normal star formation \citep{Harrison2012, Lanzuisi2015}.

We identify AGN in our three populations using the `BPT' emission line diagnostic diagram of \citet{Baldwin1981}. The BPT diagram identifies the dominant source of ionization in a galaxy by its position in log$_{10}$([N\textsc{ii}$_{\lambda6584}$]/H$\beta$) vs. log$_{10}$([O\textsc{iii}$_{\lambda5007}$]/H$\alpha$) space. The distributions of Red Misfits, Blue Actives and Red Passives in the emission-line sample on the BPT diagram are illustrated in Fig. \ref{fig:BPT}. The solid line of \citet[]{Kauffmann2003c}, based on the observed distribution of SDSS galaxies, provides the upper limit for galaxies whose emission is dominated by ongoing star formation (i.e. photoionization from O and B stars). The dashed line of \citet[][]{Kewley2001}, hereafter Ke01, based on the region which can reasonably be described by maximal starburst emission, provides the lower limit for galaxies whose emission is dominated by non-stellar processes. \textcolor{black}{This region above the Ke01 line, often referred to as the `AGN' region, is further subdivided into a `Seyfert' region and a `LINER' \citep[low ionization nuclear emission-line regions;][]{Heckman1980} region using the line defined in \citet{Schawinski2007}.}

Galaxies in the Seyfert region are objects whose emission is dominated by a luminous AGN. The LINER region, however, is more controversial. A  variety of physical processes can be invoked as an explanation of LINER emission, from low-luminosity, radiatively-inefficient AGN \citep[e.g.][]{Ho1993,Ho1997a,Kewley2006}, fast radiative shocks \citep[e.g.][]{Dopita1995,Dopita2015}, or photoionization from young stars \citep[e.g.][]{Filippenko1992,Shields1992}. See \citet{Ho2008} for a review of suspected LINER sources. There is mounting observational evidence of LINER-like emission from evolved stellar populations (mainly post-AGB stars) in optically red galaxies \citep[e.g.][]{Binette1994, Stasinska2008, CidFernandes2011, Yan2012}. Recent spatially-resolved observations of LINERs support this picture, showing that the LINER emission can be spread over kpc scales instead of confined to the nucleus \citep{Sarzi2006, Sarzi2010, Belfiore2016}.

From Fig. \ref{fig:BPT} it is clear that the vast majority of emission-line Blue Active galaxies reside in the star-forming region, although there is a small tail in the Seyfert+LINER region above the Ke01 line. Conversely, there are very few ($\sim$5 per cent) emission-line Red Passive galaxies in the star-forming region. Rather, the emission-line Red Passive distribution peaks in the composite and LINER regions with a significant tail into the Seyfert region. Finally, emission-line Red Misfits are concentrated in the star-forming and composite regions but have a significant tail ($\sim$13 per cent) into the Seyfert+LINER region. Only $\sim$4\% of emission-line Red Misfits lie in the LINER region, therefore the contribution of LINER-like emission to the Red Misfit population is minimal, no matter the excitation mechanism invoked.

Fig. \ref{fig:agn}a shows the fraction of galaxies identified as hosting an AGN (i.e. fraction of galaxies lying above the Ke01 line) in the emission-line sample binned by stellar mass. Across the entire stellar mass range of our emission-line sample, emission-line Red Passives are the most likely of the three populations to be located above the Ke01 line on a BPT diagram, especially toward higher stellar masses. Emission-line Red Misfits are significantly more likely to host an AGN than Blue Actives, especially at low stellar mass where there are virtually no AGN hosted by Blue Active emission-line galaxies. 

We stress that the emission-line sample is not representative of the full sample. The vast majority of Red Passives do not exhibit strong emission and are underrepresented in the emission-line sample as described in Section \ref{emission} (see also Table \ref{Table 1}). The Red Passives that are included in the emission-line sample are going to be the ones with significant non-stellar emission, hence the high f$_{AGN}$ in Fig. \ref{fig:agn}a. Conversely, the significant emission from star formation means nearly all Blue Actives are included in the emission-line sample and consequently a relatively small fraction of them will host an AGN. 

Fig. \ref{fig:agn}b shows the fraction of galaxies hosting AGN in the full sample, i.e. the fraction of galaxies in the full sample that both satisfy our emission-line criteria outlined in Section \ref{emission} and lie above the Ke01 line on a BPT diagram. At all stellar masses, we find Red Misfits are more likely to host an AGN than Blue Actives or Red Passives in the full sample. The f$_{AGN}$ enhancement in Red Misfits is particularly strong towards low stellar mass, where the fraction of Red Passive galaxies hosting AGN is low and the fraction of Blue Actives hosting AGN is nearly zero. \textcolor{black}{We stress that this f$_{AGN}$ enhancement is dominated by the sizeable population of emission-line Red Misfits in the Seyfert region of the BPT diagram. There is no significant enhancement in the fraction of Red Misfits identified as LINERs in our full sample when compared to Red Passives.}

Although Red Passives are significantly less likely to host a BPT-identified AGN when compared to Red Misfits across the entire stellar mass range, Red Passives are still the most common hosts of AGN in our full sample (particularly at high stellar mass) due to the sheer size of the population (see Table \ref{Table 1}). However, below a stellar mass of $\sim$10$^{10.3}$ M$_{\sun}$, Red Misfits are the most common hosts of BPT-identified AGN in the local Universe.

\subsection{Environmental Trends}
\label{Environment}

\begin{figure*}
	\centering
	\includegraphics[width=\textwidth]{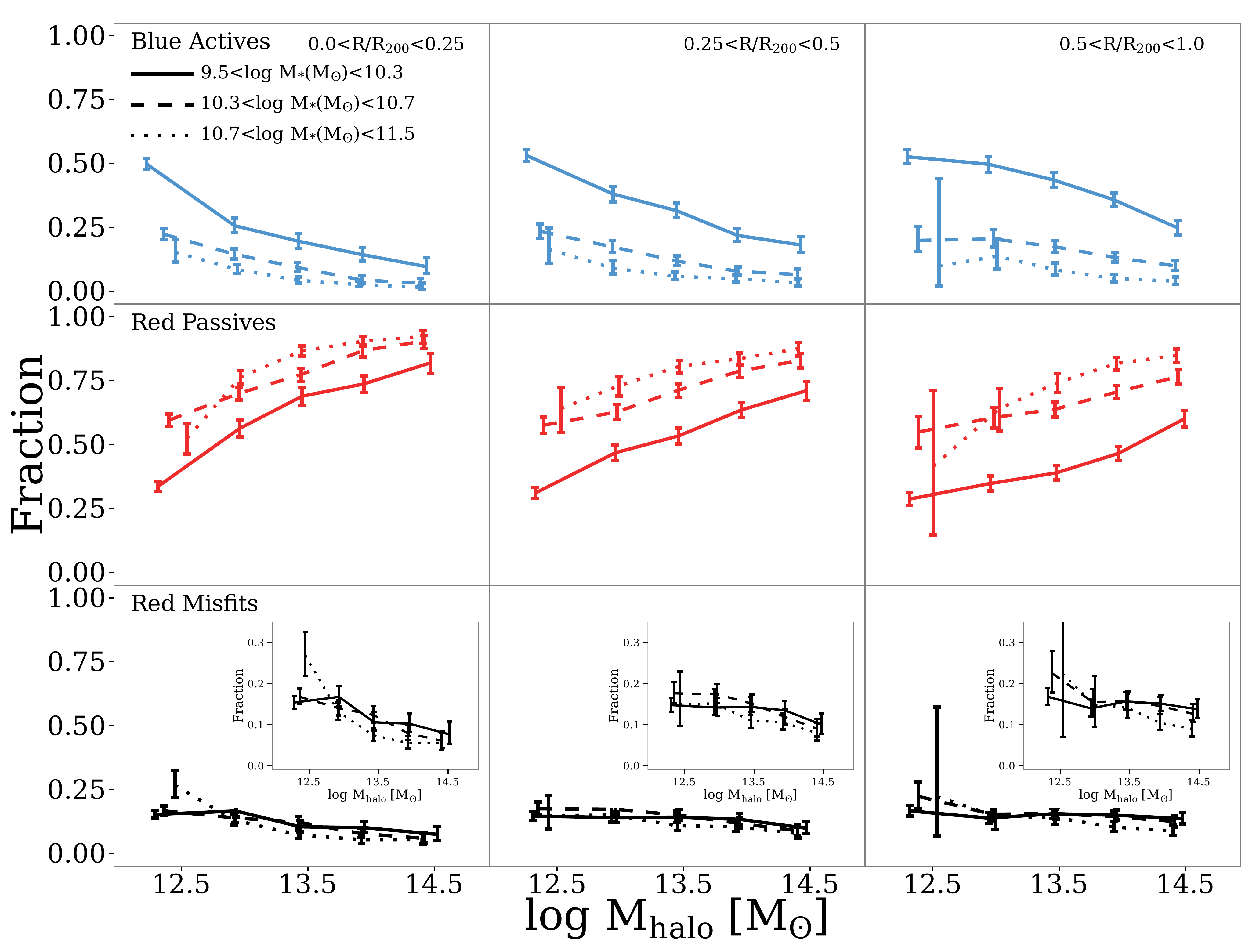}	
	\caption{V$_{max}$-weighted satellite fractions of Blue Actives (top), Red Passives (middle) and Red Misfits (bottom) in our group sample against group halo mass. Fractions are shown for galaxies in the inner third (left column), middle third (middle column) and outer third (right column) of the R/R$_{200}$ distribution. Results are also shown in three bins of stellar mass as different line styles corresponding to the upper, middle and lower thirds of the stellar mass distribution. Insets zoom in on the Red Misfit results. Error bars are 99\% confidence intervals generated by the beta distribution as outlined by \citet{Cameron2011}.}  
	\label{fig:FracMhaloDist}
\end{figure*}

\begin{figure*}
	\centering
	\includegraphics[width=\textwidth]{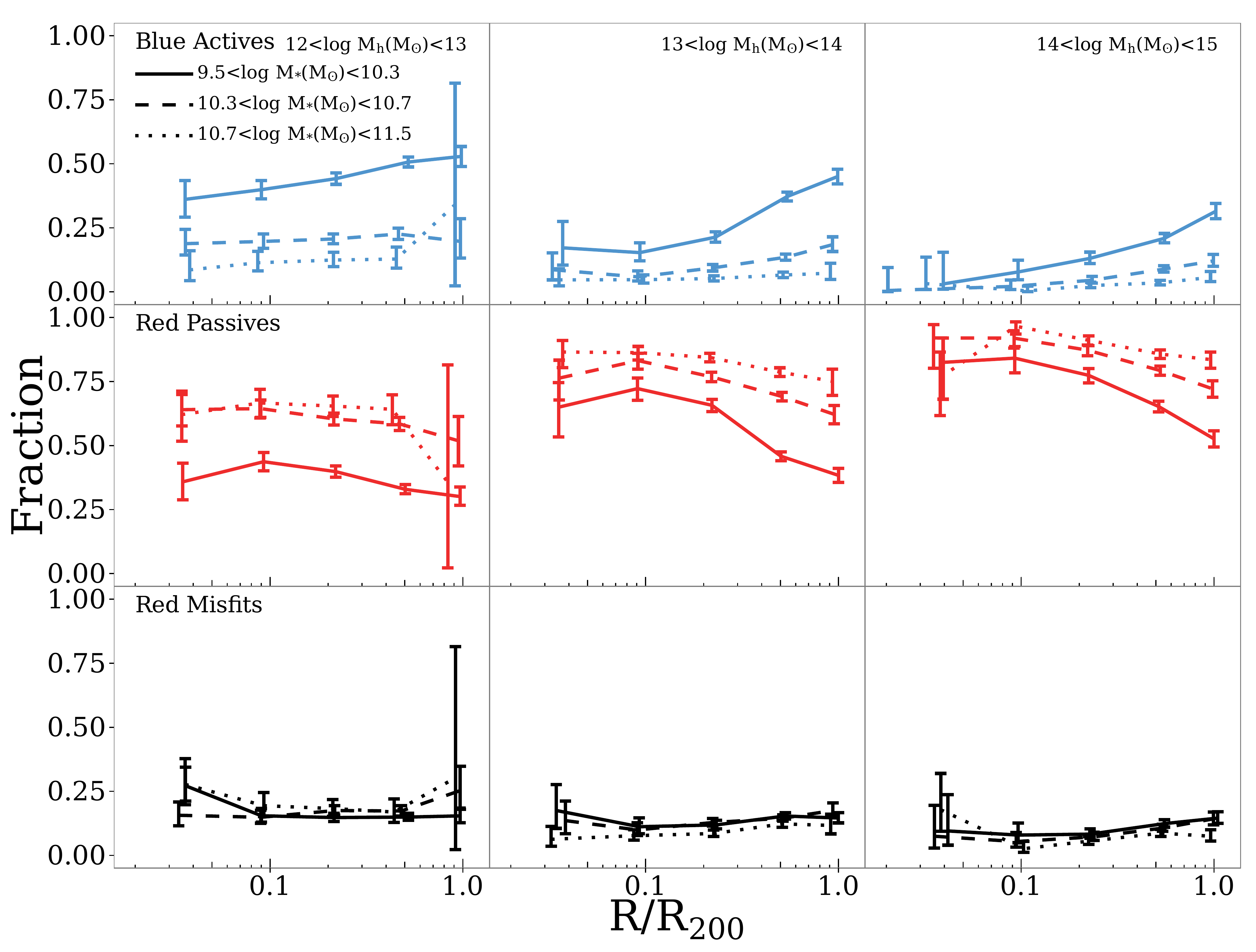}
	\caption{Same as Fig. \ref{fig:FracMhaloDist} except we plot fractions against R/R$_{200}$ and bin by group halo mass in the columns.}  
	\label{fig:FracDistMhalo}
\end{figure*}

\begin{figure*}
	\centering
	\includegraphics[width=\textwidth]{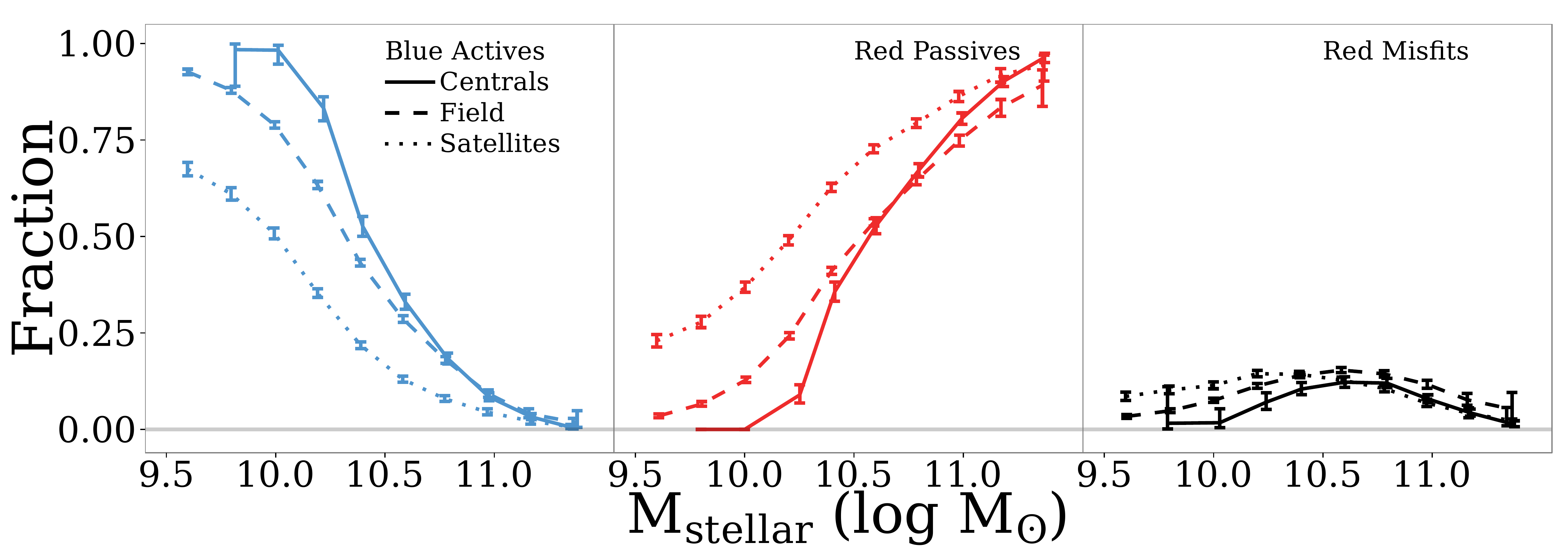}
	\caption{V$_{max}$-weighted fractions of Red Misfit, Blue Active and Red Passive galaxies with stellar mass in our isolated field sample, the sample of centrals in our group catalogue and the sample of satellites in our group catalogue. The fraction of Red Misfits depends only weakly on stellar mass and environment. Error bars are 99\% confidence intervals generated by the beta distribution as outlined by \citet{Cameron2011}. }
	\label{fig:FieldCent}
\end{figure*}

Having explored the stellar masses, morphologies, dust content and AGN abundance of Red Misfits compared to Blue Actives and Red Passives, we now consider how the proportions of these populations change with environment. We characterize environment by both group halo mass, given in the Y07 catalogue, and group-centric distance scaled by the characteristic radius R$_{200}$ described in Section \ref{group}. Given that star formation trends with environment are dominated by the changing fraction of passive galaxies rather than the changing sSFR of star-forming galaxies themselves \citep[e.g.][]{McGee2011, Wijesinghe2012, Schaefer2017}, exploring how population proportions vary with environment can help constrain the physical processes responsible for quenching.

We show the V$_{max}$-weighted relative fractions of each population against group halo mass in Fig. \ref{fig:FracMhaloDist}. \textcolor{black}{Results are shown for samples binned by stellar mass and group-centric distance.} The Blue Active population fraction decreases with increasing group halo mass with a stronger halo mass dependence at low stellar mass. Correspondingly, the Red Passive fraction increases with halo mass. The Red Misfit dependence on halo mass is much weaker, but the inset shows that the fraction of Red Misfits decreases weakly with halo mass in all bins of stellar mass and group-centric distance with little dependence on stellar mass.

To see radial trends more clearly, Fig. \ref{fig:FracDistMhalo} shows population fractions plotted against group-centric distance and binned in columns by group halo mass. The fraction of Blue Actives (Red Passives) increases (decreases) with increasing group-centric distance with a stronger dependence at low stellar mass, though generally the \textcolor{black}{population trends with group-centric distance} are weaker than the trends with halo mass.

Population fractions with stellar mass in isolated, central and satellite samples are \textcolor{black}{compared} in Fig. \ref{fig:FieldCent}. Blue Active and Red Passive fractions depend most strongly on stellar mass in our sample of centrals and most weakly in our sample of satellites. Red Misfit fractions, however, are remarkably flat with both stellar mass and environment. With the exception of very low stellar mass, where the fraction of Red Misfit centrals drops to zero, there is an $\sim$11 per cent chance of a galaxy in a given sample being a Red Misfit, independent of its stellar mass or whether it is a central, satellite or isolated galaxy. We note that the overwhelming majority of low-stellar mass centrals are centrals of galaxy pairs or triplets rather than richer groups and in very low mass groups there is a risk of misidentifying the central galaxy \citep[e.g.][]{Peng2012, Knobel2013}. 

\section{Discussion}

In the previous sections we have shown that there exists a small but significant population of red and star-forming galaxies in the Sloan Digital Sky Survey. These `Red Misfit' galaxies exhibit intermediate morphologies relative to the more common Blue Active and Red Passive populations, show a slight excess of \textcolor{black}{reddening} due to intrinsic dust, and have an enhanced probability of hosting an AGN, This population represents $\sim$11 per cent of $z<0.1$ galaxies, varying only weakly with stellar mass, halo mass, and group-centric radius. In this section we interpret these results to try and elucidate the physical origin of Red Misfits and \textcolor{black}{place Red Misfits in the context of other outlier galaxy populations in the local Universe.}

\subsection{Red Misfits as Having Poorly-Constrained Star Formation}

Red Misfits are defined based on their colours and sSFRs. \textcolor{black}{Star formation rates for SDSS galaxies often rely on H$\alpha$ and/or [O\textsc{II}] emission line luminosities \citep[see reviews by][]{Kennicutt1998, Kennicutt2012}, so the SFRs and sSFRs for AGN-contaminated and low S/N galaxies can be poorly constrained \citep[e.g.][]{Hopkins2003, Brinchmann2004}}. The overestimation of Red Misfit sSFRs due to AGN contamination or low S/N emission lines is a valid concern, especially considering the preponderance of AGN in Red Misfits. We use the updated, $GALEX$ near- and far- UV-based sSFRs of \citet{Salim2007}. Whereas AGN can contaminate H$\alpha$, the contribution to the UV continuum from AGN is low \citep{Salim2007,Kauffmann2007}. 

Despite these updated prescriptions, typical uncertainties on sSFR estimates are significant, especially for redder galaxies with weak or undetected emission lines. The median 1$\sigma$ scatter in sSFR of Red Passives is 0.8 dex (although it decreases towards higher sSFR). To ensure that `smearing' of the Red Passive population due to measurement uncertainties is not responsible for the Red Misfit population, we test whether our qualitative results vary with more conservative cuts on sSFR. Instead of using the median sSFR for each galaxy we instead use the full sSFR probability distributions and select star-forming galaxies as those with $\geq$84\%  probability of lying above our sSFR cut of 10$^{-10.8}$ yr$^{-1}$. In doing so, our Red Misfit sample is reduced to $\sim$50\% its original size but the trends presented in this work remain.

We can also focus on the emission line sample and identify galaxies as star-forming using a BPT diagnostic (see Fig. \ref{fig:BPT}). This selection results in a much smaller sample of Red Misfits, but their properties (e.g. their nearly constant fraction with stellar mass, intermediate morphologies, increased dust and gas-phase metallicity) remain the same as those identified with an sSFR selection. We can therefore not attribute the Red Misfit population to uncertain sSFR measurements of quiescent, red galaxies.

\subsection{Red Misfits as a Combination of Blue Actives and Red Passives} 

In Section \ref{Dust} we show that Red Misfits are not entirely dusty star-forming galaxies, however it is possible that Red Misfits could be a mix of dusty Blue Actives and Red Passives with overestimated sSFRs. Indeed, inclination-induced dust reddening significantly inflates the population of Red Misfits if inclination-corrected colours are not used (see Appendix \ref{App:Maller}). However, if the Red Misfits that remain after inclination correction are dominated by Blue Actives or Red Passives, their properties should closely match one of the two populations. Red Misfits and Blue Actives are similar in their Balmer decrements (see Fig. \ref{fig:Balmer}), mean stellar ages and gas-phase metallicities (see Fig. \ref{fig:Dn4000}). On the other hand, the morphologies and AGN fractions of Red Misfits are typical of neither Blue Actives nor Red Passives.

Previous authors \citep[e.g.][]{Gallazzi2008, Mahajan2009} find bimodality in properties of red star-forming galaxies, which suggests that the population could be a mix of two distinct populations. We stress that we do not see evidence of multiple components for any property of our Red Misfits -- they appear to be a single population. There could be some contamination of the Red Misfits by Blue Actives and Red Passives with uncertainties in sSFR or colour. Given uncertainties, at most 5.7 per cent of Blue Actives and 11 per cent of Red Passives could be classified as Red Misfits, which cannot account for the entire population of Red Misfits. Therefore while there may be some contamination of our Red Misfit sample by Blue Actives and Red Passives, it is impossible to combine these populations in a way that reproduces the properties of Red Misfits. Even under the most conservative sSFR cuts and colour corrections, a population of red and star-forming galaxies persists.

\subsection{Red Misfits as a Distinct Population} \label{distinct}

\begin{figure*}
	\centering
	\includegraphics[width=0.49\textwidth]{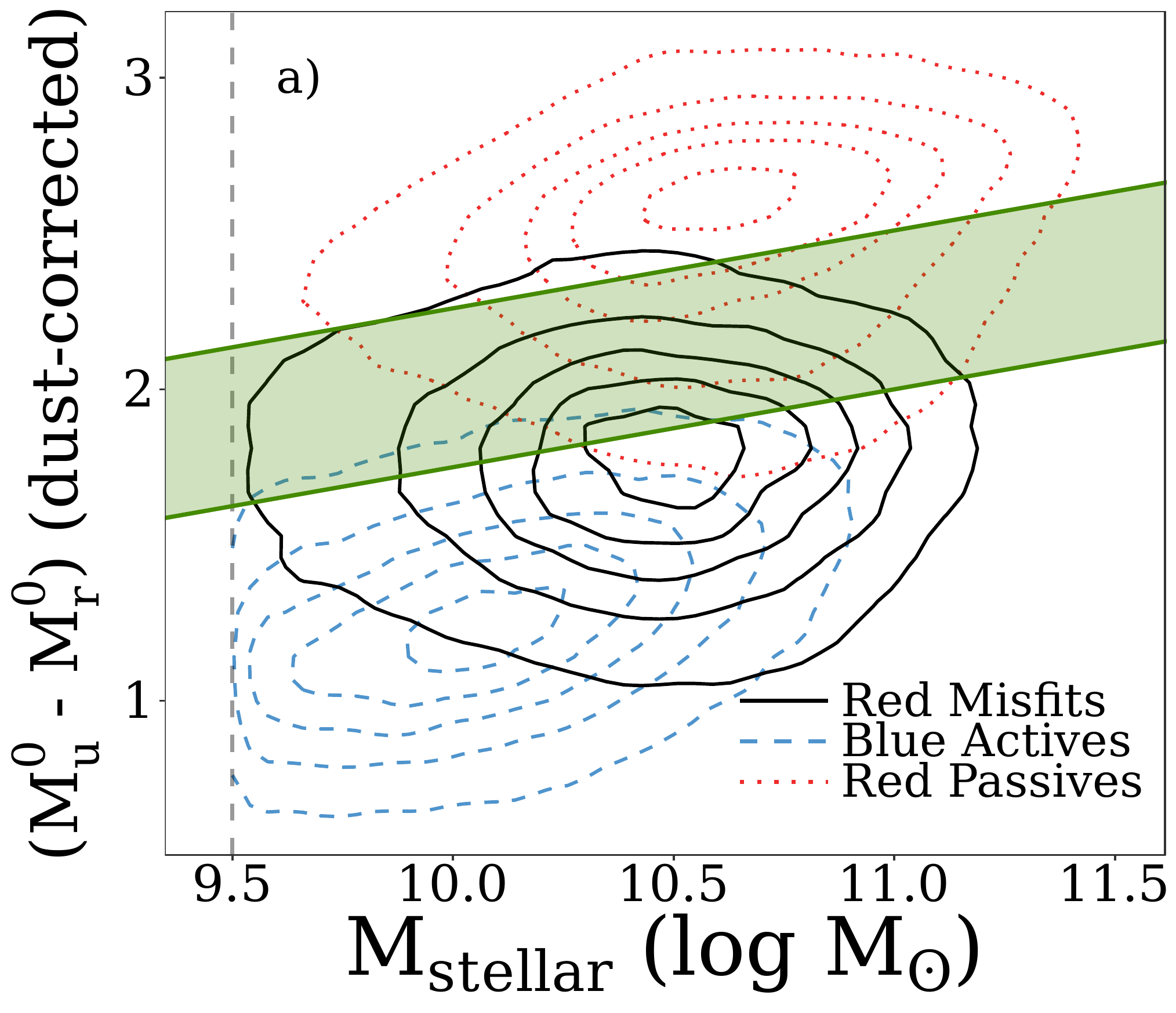}	
	\includegraphics[width=0.49\textwidth]{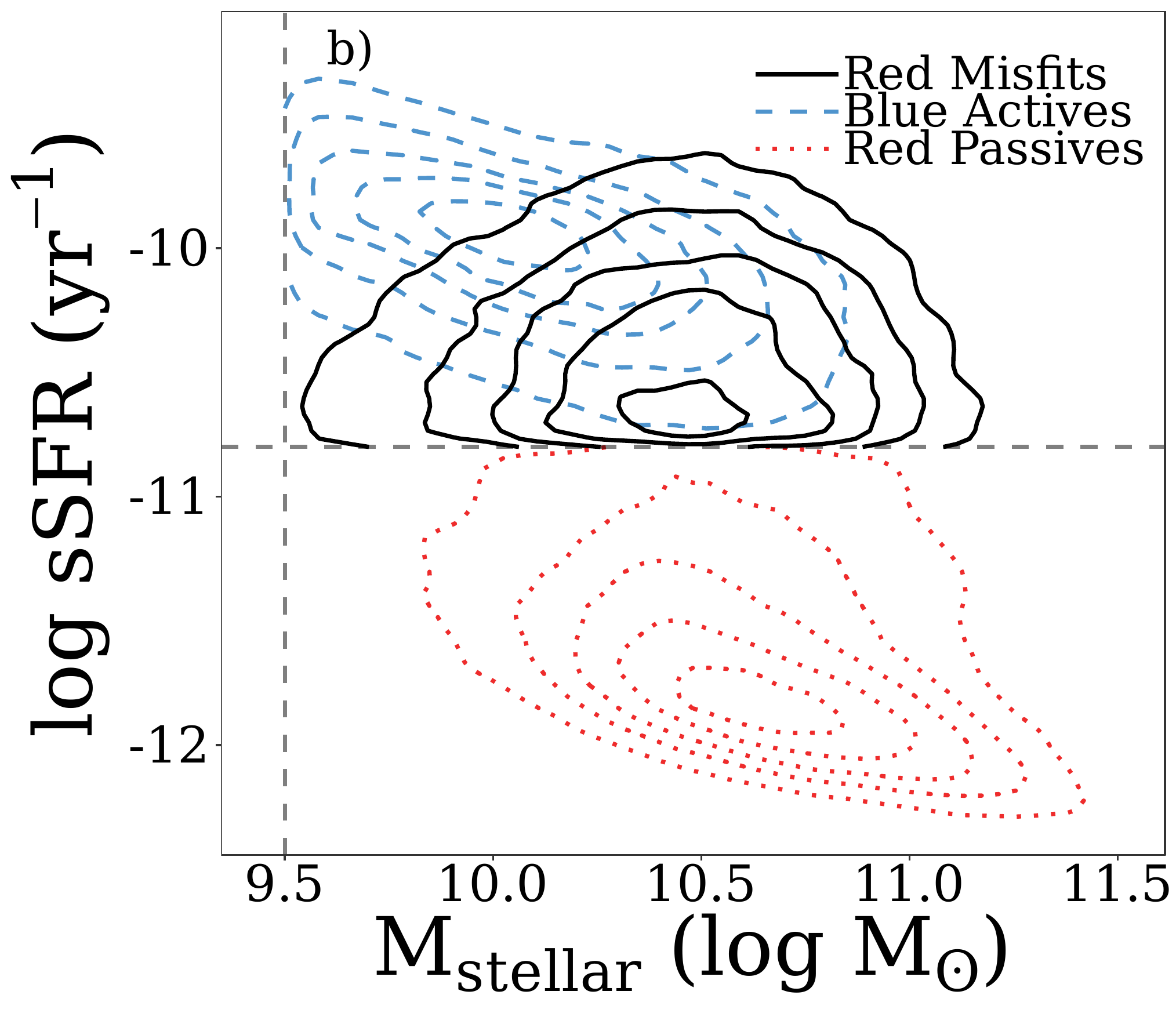}	
	\caption{\textcolor{black}{\textit{a):} Rest-frame k-corrected $u-r$ colour against stellar mass for Red Misfits, Blue Actives and Red Passives. Colours are \textit{not} corrected for inclination but are corrected for dust reddening using the stellar continuum $E(B-V)$ as measured by \citet{Oh2011}. Shaded green region shows the green valley as defined by \citet{Schawinski2014}. \textit{b)}: sSFR against stellar mass for Red Misfits, Blue Actives and Red Passives. Contours in both plots encompass 10\%, 30\%, 50\%, 70\% and 90\% of the unweighted distributions. Dashed lines indicate our M$_{stellar}$=10$^{9.5}M_\odot$ and sSFR=10$^{-10.8}$ yr$^{-1}$ cuts.}}
	\label{fig:valley}
\end{figure*}

Given that the morphologies, AGN abundances, dust content and environmental trends of SDSS Red Misfits are inconsistent with being dusty/edge-on Blue Actives or Red Passives with overestimated sSFRs (or a combination thereof) and there is little evidence they have particularly ill-constrained sSFRs, we suggest instead that Red Misfits are in fact their own separate population of galaxies in the local Universe independent from the Blue Active and Red Passive populations. Here we explore the possible physical origins of Red Misfits, their connection to other known populations and how they fit into the broad picture of galaxy evolution.

\subsubsection{Connection with the Green Valley}

Red Misfits exhibit many intermediate properties which indicate they might be a transition population between the blue star-forming cloud and the red sequence. They may be similar to previously identified intermediate populations such as the green valley -- the sparsely-populated region between the red sequence and the blue cloud \citep[e.g.][]{Bell2004, Faber2007, Martin2007, Wyder2007}. \textcolor{black}{In Fig. \ref{fig:valley}a we show the distribution of Red Misfits, Blue Actives and Red Passives in $u-r$ vs. $M_{stellar}$. To better enable comparison to previous work we use z=0.0 k-corrected magnitudes uncorrected for inclination but corrected for dust extinction in the stellar continuum using the \citet{Calzetti2000} reddening law and the $E(B-V)$ measurements of \citet{Oh2011} based on the \texttt{GANDALF} code \citep{Cappellari2004, Sarzi2006}. The shaded green region in Fig. \ref{fig:valley}a shows the green valley as defined in Fig. 4 of \citet{Schawinski2014}. Although a sizable proportion of Red Misfits (34.5 per cent) lie in the green valley, a significant number of Red Passives lie in the green valley as well (30 per cent).}

\textcolor{black}{It is clear in Fig. \ref{fig:valley}b} that sSFRs and stellar masses of Blue Actives are tightly correlated, forming the \textcolor{black}{specific} star-forming main sequence found in samples of star-forming galaxies \citep[e.g.][]{Brinchmann2004, Noeske2007b, Elbaz2007}. \textcolor{black}{The Red Misfit distribution, however, does not exhibit this correlation and the locus lies predominantly below the main sequence of star formation.}

The Red Misfit distributions in Fig. \ref{fig:valley} therefore support a scenario in which Red Misfits are in fact transition galaxies similar to green valley galaxies in the midst of being quenched gradually on their way to the red sequence. Star formation is suppressed but not entirely zero in these galaxies, therefore their stellar mass continues to increase as they fall off the specific star-forming main sequence and gradually redden in colour, explaining the older mean stellar age and enhanced metallicity when compared to Blue Actives. 

This scheme of gradual quenching is qualitatively similar to the late-type quenching model of \citet{Schawinski2014} \textcolor{black}{and a follow-up investigation \citep{Smethurst2015} finding that the green valley is dominated by disk and intermediate-morphology galaxies evolving slowly.} An important distinction, however, is that \citet{Schawinski2014} infer a significant environmental contribution to this quenching model as they find a dearth of Galaxy Zoo-selected late-type green valley galaxies in halos below M$_{halo}$=10$^{12}$ h$^{-1}$M$_{\odot}$. The proportion of Red Misfits, however, is \textcolor{black}{constant with increasing} M$_{halo}$.

It is this (near) equal abundance of Red Misfits in all environments that suggests the processes that govern/create them are driven \textcolor{black}{primarily} by internal evolution. The prevalence of AGN in Red Misfits suggests they play a role, either as a mechanism of internal quenching or a byproduct of it. AGN feedback can suppress gas cooling on 1-2 Gyr time-scales \citep[e.g.][]{Croton2006}, inhibiting star formation after the cool gas supply is exhausted. The more dominant bulges of Red Misfits when compared to Blue Actives may indicate more massive central black holes \citep{Magorrian1998,Haring2004} and thus stronger past feedback. AGN have been found to occupy similar regions in parameter space as Red Misfits, i.e. in the green valley \citep[e.g.][]{Nandra2007,Silverman2008, Schawinski2010, Schawinski2014} and below the massive end of the specific star-forming main sequence \citep{Salim2007}. The intermediate morphologies of Red Misfits are also consistent with the green valley \citep{Mendez2011,Smethurst2015}.

AGN feedback is not the only internal mechanism capable of quenching star formation. \textcolor{black}{The deep potential well of a significant bulge created by internal processes or minor mergers combined with the declining self-gravity of the cool gas as the gas supply dwindles can prevent the fragmentation of the disk and thus suppress star formation \citep{Martig2009, Fang2013}. Additionally,} galaxy bars are efficient at driving gas inwards \citep[e.g.][]{Athanassoula1992, Knapen1995}, exhausting cold gas supply by spurring central star formation \citep[e.g.][]{Martin1997, Zurita2004, Sheth2005}, building a pseudobulge \citep[e.g.][]{Kormendy2004,Athanassoula2005}. \textcolor{black}{Bars may also explain the enhanced abundance of AGN in Red Misfits, however, results investigating the connection between the presence of a bar and AGN activity have been inconclusive, with some groups establishing a link \citep[e.g.][]{Knapen2000,Laine2002,Galloway2015} and others finding no causal connection \citep[e.g.][]{Ho1997b, Cisternas2013, Cisternas2015, Cheung2015}.} 

\subsubsection{Connection with S0s/S0 Projenitors}

Previous work studying anomalously red galaxies \citep[e.g.][]{Goto2003,Wolf2009,Koyama2011} have suggested that they are the projenitors of the S0 population. S0s have also been discussed as a transition population in the context of transforming infalling blue galaxies \citep[e.g.][]{Poggianti1999, Kodama2001, Wilman2012}.

At a glance there are many similarities between Red Misfits and S0s: their red colours \citep{Roberts1994}, enhanced sSFR over elliptical galaxies \citep[e.g.][]{Amblard2014}, overlap with the green valley \citep{Salim2012} intermediate bulge-total ratio relative to early-type and late-types \citep[e.g.][]{Simien1986}, enhanced AGN fraction over ellipticals \citep[e.g.][]{Schawinski2007, Wilman2012, Amblard2014} and an enhanced dust mass relative to ellipticals as well \citep[][]{Amblard2014}. \textcolor{black}{While S0s can be found in a wide range of environments \citep[e.g.][]{Dressler1980, Dressler1997, VandenBergh2009, Wilman2009}, the fact that they are most abundant in clusters \citep{Dressler1980} and that the S0 fraction depends on both halo mass and stellar mass \citep{Wilman2012} are not consistent with Red Misfits.} So while the Red Misfit population may include some S0 projenitors, it cannot explain the whole population.

\subsubsection{Connection with Red Spirals}
The red colours and early-disc morphologies of Red Misfits are reminiscent of the significant \textcolor{black}{population of `red spiral' galaxies} in the Universe \citep{VandenBergh1976, Dressler1999, Poggianti1999}. Indeed, some authors investigating a red star-forming population \citep[e.g.][]{Gallazzi2008, Wolf2009, Crossett2014} treat red spirals and star-forming red galaxies as largely the same \textcolor{black}{population}. Red spirals are often referred to as \textcolor{black}{quiescent} \citep[e.g.][]{Poggianti1999, Goto2003} yet many can still exhibit significant star formation \citep{Cortese2012}. \textcolor{black}{Indeed, 35 per cent of the red spirals in the catalogue of \citet{Masters2010} are classified as Red Misfits by our sSFR and inclination-corrected colour definitions (see Section \ref{Classification}).} Red Misfits and red spirals both show intermediate star formation rates \citep{Tojeiro2013} and enhanced AGN fractions \citep{Masters2010}, however red spirals show dust and environment properties \citep[e.g.][]{Goto2003, Skibba2009, Bamford2009, Masters2010, Rowlands2012, Tojeiro2013} which are distinct from Red Misfits. \textcolor{black}{Additionally, the enhanced AGN fraction among red spirals in \citet{Masters2010} is driven primarily by galaxies dominated by LINER-like emission, which is not the case for Red Misfits (see Fig. \ref{fig:BPT})}. These red spirals have been suggested as a projenitor of S0s \citep{Goto2003, Wolf2009} and as a evolutionary stage on the way to the red sequence \citep{Moran2007, Bundy2010, Tojeiro2013}. \textcolor{black}{It is possible that red spirals with large bulges (note that not all selections allow for the inclusion of early-type spirals) represent Red Misfits which have finished quenching.} 

\section{Summary \& Conclusions}

We investigate a population of red, star-forming galaxies (Red Misfits) in the Sloan Digital Sky Survey Data Release 7. We identify them as star-forming based on their MPA-JHU specific star formation rates and red using NYU-VAGC $g-r$ colours, k-corrected to $z=0.1$ and corrected for attenuation due to inclination. This population represents $\sim$11 per cent of galaxies in the local Universe. Selecting galaxy populations using \textit{both} sSFR and colour allows more robust identification of intermediate or transition galaxies than using only one or the other. The use of inclination-corrected colours reduces contamination of intermediate populations by highly-inclined galaxies, allowing more physically meaningful classification of galaxy populations. 

To elucidate the physical origin of Red Misfits and their role in galaxy evolution, we study the properties and halo occupation statistics of Red Misfits, comparing them to Blue Active and Red Passives and finding the following:

\begin{enumerate}
	\item The fraction of Red Misfits is nearly \textcolor{black}{constant with increasing} stellar mass (Fig. \ref{fig:mass}). This is in stark contrast to Blue Actives and Red Passives, which dominate demographics in low and high stellar mass samples respectively.
	\item \textcolor{black}{Red Misfits are most often classified as having intermediate morphologies with both the machine learning visual classifications of \citet{Huertas2011} and the S\'{e}rsic indices of \citet{Simard2011}.} Red Misfits are not universally early-type or late-type but rather exhibit a range of morphologies and prefer intermediate morphologies. 
	\item At fixed stellar mass, Red Misfits show slightly larger levels of intrinsic dust-reddening than Blue Actives as measured by the Balmer decrement (Fig. \ref{fig:BPT}). This excess, however, is not enough to attribute the colours of Red Misfits entirely to dust. Rather, their colours are \textcolor{black}{likely} due to contributions from increased dust \textcolor{black}{and slightly older stellar populations as compared to Blue Actives.} 
	\item Overall, Red Misfits in our smaller emission-line sample are twice as likely to be \textcolor{black}{identified as hosting an AGN by a BPT diagnostic} than Red Passives and 5.5 times more likely to host \textcolor{black}{an AGN} than Blue Actives (Fig. \ref{fig:agn}). This enhanced AGN likelihood persists at fixed stellar mass and is especially strong at low stellar mass.
	\item Red Misfits exhibit only very weak environmental trends. Their proportion declines slightly with increasing host halo mass but is flat with increasing group-centric separation (Fig. \ref{fig:FracMhaloDist}, \ref{fig:FracDistMhalo}). Blue Actives and Red Passives, by comparison, exhibit much stronger environmental trends in our group sample.  Additionally, Red Misfits are found in equal proportion in our isolated and group samples. Conversely, Blue Actives are more abundant in the isolated  sample and Red Passives more abundant in the group sample when compared to our total sample. 
	\item Red Misfits overlap \textcolor{black}{and share some properties} with previously identified intermediate populations (e.g. green valley galaxies, S0s) however they show different trends with environment. 
\end{enumerate}

We interpret these results as evidence that Red Misfits are a transition population of galaxies between the blue cloud and the red sequence. These galaxies are being gradually quenched; their sSFRs slowly moving off the star-forming main sequence as they continue to grow in stellar mass and their stellar population ages. Red Misfits' agnosticism towards environment indicates that this gradual quenching is primarily driven by internal processes.

\section*{Acknowledgements}
\textcolor{black}{We thank the anonymous referee whose thoughtful comments have improved this manuscript.} FAE acknowledges support from an Ontario Graduate Scholarship. LCP and IDR acknowledge support from the Natural Sciences and Engineering Research Council of Canada. \textcolor{black}{This work made us of many open-source software packages such as \texttt{ggplot2} \citep{Wickham2009}, \texttt{MASS} \citep{Venables2002}, and \texttt{TOPCAT} \citep{Taylor2005}.}

Funding for the SDSS and SDSS-II has been provided by the Alfred P. Sloan Foundation, the Participating Institutions, the National Science Foundation, the U.S. Department of Energy, the National Aeronautics and Space Administration, the Japanese Monbukagakusho, the Max Planck Society, and the Higher Education Funding Council for England. The SDSS Web Site is http://www.sdss.org/. The SDSS is managed by the Astrophysical Research Consortium for the Participating Institutions. The Participating Institutions are the American Museum of Natural History, Astrophysical Institute Potsdam, University of Basel, University of Cambridge, Case Western Reserve University, University of Chicago, Drexel University, Fermilab, the Institute for Advanced Study, the Japan Participation Group, Johns Hopkins University, the Joint Institute for Nuclear Astrophysics, the Kavli Institute for Particle Astrophysics and Cosmology, the Korean Scientist Group, the Chinese Academy of Sciences (LAMOST), Los Alamos National Laboratory, the Max-Planck-Institute for Astronomy (MPIA), the Max-Planck-Institute for Astrophysics (MPA), New Mexico State University, Ohio State University, University of Pittsburgh, University of Portsmouth, Princeton University, the United States Naval Observatory, and the University of Washington.

\bibliographystyle{mnras}
\bibliography{Fraser_Evans_Bib}

\appendix

\section{Determining Intrinsic Galaxy Colours} \label{App:Maller}
\begin{figure*}
		\renewenvironment{figure}[1][]{\ignorespaces}{\unskip}
		\includegraphics[width=0.49\textwidth]{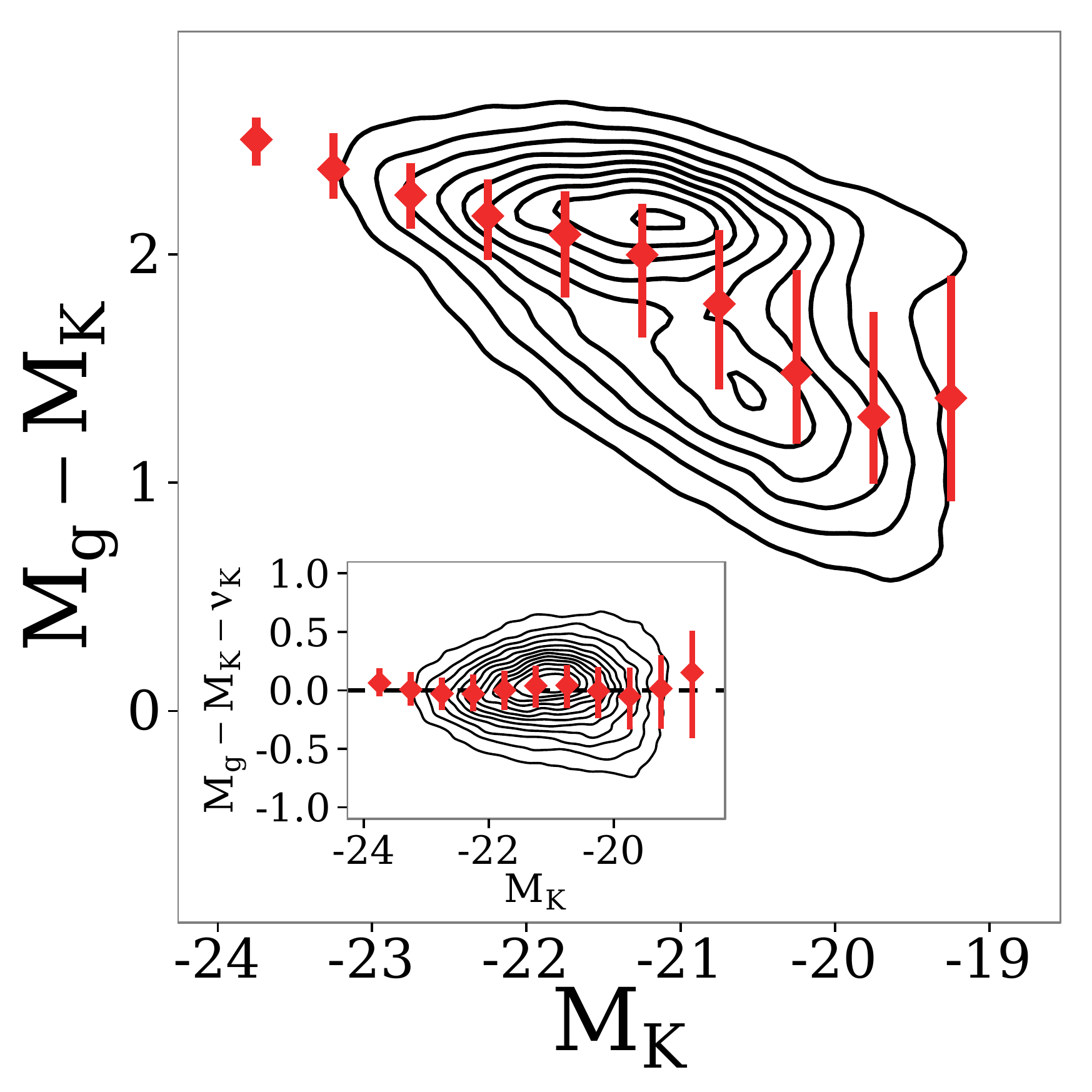}
		\includegraphics[width=0.49\textwidth]{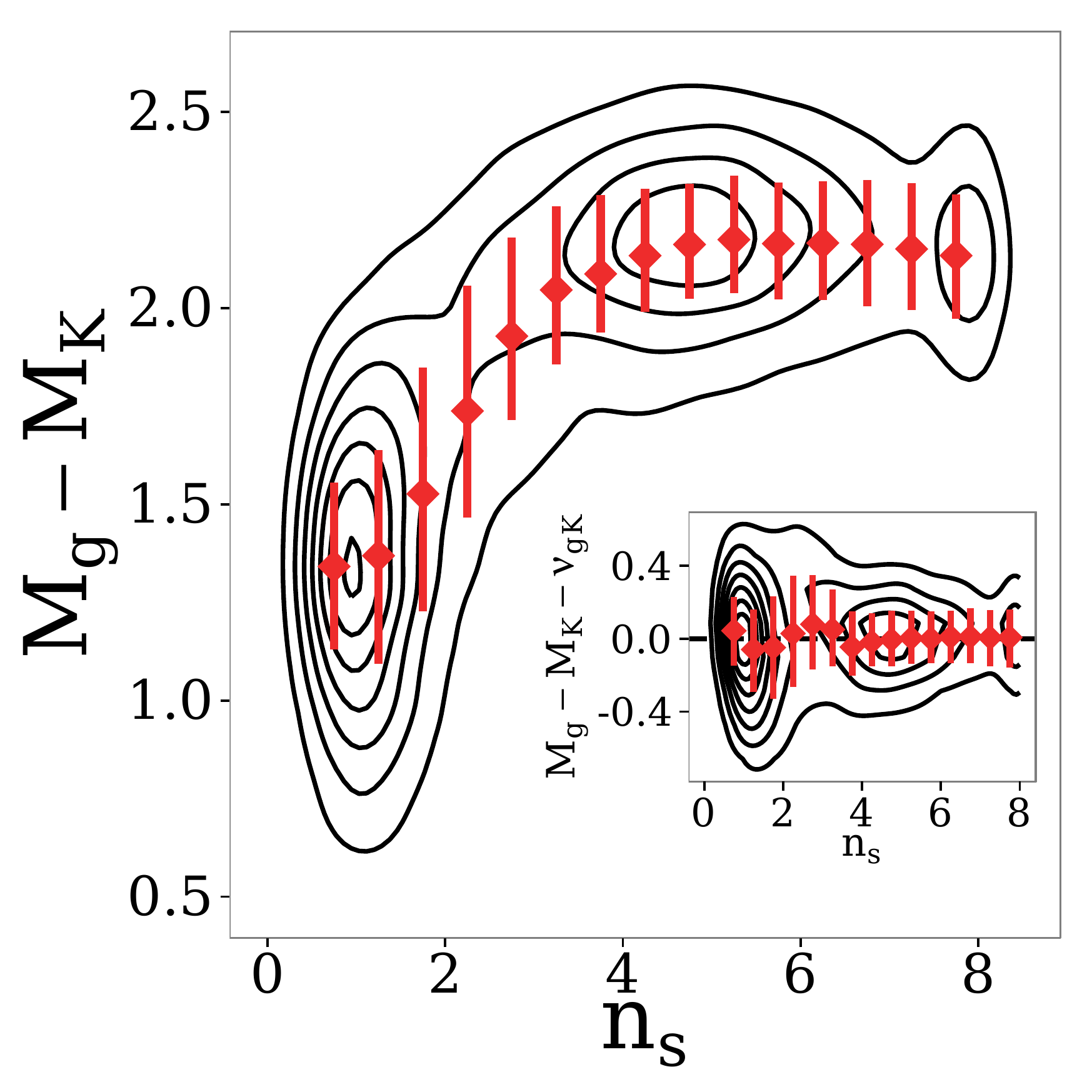}
	\caption{M$_g$ - M$_K$ colour of face-on (b/a>0.85) galaxies against M$_K$ (left) and S\'{e}rsic index (right). Points show median value and interquartile range in each bin. Insets show residuals between M$_g$ - M$_K$ colour and Eq. \ref{ColFit}.} 
	\label{fig:FaceOnColour}
\end{figure*}

The classification of the galaxies in our catalogue into different populations depends in part on observed \textcolor{black}{$g$-$r$} colour, a property known to vary with inclination \citep[e.g.][]{Tully1998}. Dust \textcolor{black}{attenuation} through the disc of a galaxy will cause highly-inclined galaxies to appear redder than identical galaxies viewed face-on. It is therefore possible that many galaxies identified as being red in colour are highly-inclined blue galaxies. We correct out the inclination dependence of optical colour to obtain intrinsic colours. Since inclination angles provided by \citet{Simard2011} are reliable only for disc-dominated galaxies, we opt to use axis ratios provided by \citet{Simard2011} as our inclination proxy to calculate intrinsic colours for our full sample.

It is important to note that a galaxy's colour does not depend on inclination alone, nor is colour the only property to vary with inclination. Any colour dependence on inclination may be primarily due to dependence on a separate property that is also varying with inclination. To isolate the attenuating effect of inclination it is crucial that we compare galaxies at different inclinations that are as intrinsically identical as possible. This is typically done by matching samples by a property known not to vary with inclination, such as \textit{K} band magnitude \citep{Bell2001}.

Since inclination-dependent attenuation affects both the $g$ and $r$ bands, comparing $g-r$ colours of galaxies at different inclinations will couple these effects and may lead to complicated dependencies. To disentangle the effects of attenuation on M$_g$ and M$_r$ we find the correlation between observed M$_g$ - M$_K$ and M$_r$ - M$_K$ colours and inclination. Assuming inclination-induced attenuation in the K band is negligible \citep{Bell2001}, the attenuation in the $g$ and $r$ bands is then whatever correction is needed to remove the $g-K$ or $r-K$ colour dependences on inclination. The $g$ and $r$ attenuations can then be used to correct each galaxy's observed M$_{g}$-M$_{r}$ colour to what it would be if the galaxy were observed face-on. The advantage to this approach is that no prior knowledge or assumptions about dust properties or stellar populations are required to perform the correction. 

This general method of measuring inclination-induced attenuation or correcting colours has been applied by several groups \citep[e.g.][]{Tully1998,Masters2003,Shao2007,Padilla2008} but we follow closely the specific methodology of \citet{Maller2009}, who identify samples of galaxies that are intrinsically the same using galaxies closely matched in both M$_K$ and S\'{e}rsic index (n$_s$). Thus the attenuation in the \textit{g} and \textit{r} bands will be functions not only of inclination but also of M$_K$ and n$_s$. 

To correct the \textit{g/r-K} colours of inclined galaxies to their face-on colours we must first determine what these face-on colours actually are. Fig. \ref{fig:FaceOnColour} shows the M$_g$ - M$_K$ colours of face-on (b/a>0.85) galaxies against M$_K$ and n$_s$. We see that the median colour of face-on galaxies is linear in M$_K$. The dependence of colour on n$_s$ appears more complicated. We find that the median face-on colour is linear in n$_s$ for n$_s<$4 and constant in n$_s$ for n$_s \geq$4. Furthermore, we find the dependence of M$_g$ - M$_K$ on M$_K$ differs between n$_s<$4 and n$_s \geq$4 galaxies. We do not show similar analysis for M$_r$ - M$_K$ colours but the trends are the same. We fit the median galaxy colour by a function of the form:

\begin{subequations} \label{ColFit}
\begin{align}
	\nu_{g/r-K} &= \nu_{g/r,0} + \nu_{g/r,K} (M_K + 20) + \nu_{g/r,n}(n_s) &\text{for n}_s<\text{4}\\
	\nu_{g/r-K} &= \nu_{g/r,0}' + \nu_{g/r,K}' (M_K + 20) &\text{for n}_s \geq \text{4}
\end{align} 
\end{subequations}
where $\nu_{g/r-K}$ is the median $g-K$ or $r-K$ colour of a face-on galaxy and $\nu_{g/r,0}$, $\nu_{g/r,K}$ and $\nu_{g/r,n}$ are fitting parameters. Residuals to this fit are shown in insets in Fig. \ref{fig:FaceOnColour}, demonstrating that our model fits the galaxy colours well. 

\begin{figure*}
	\includegraphics[width=0.95\textwidth]{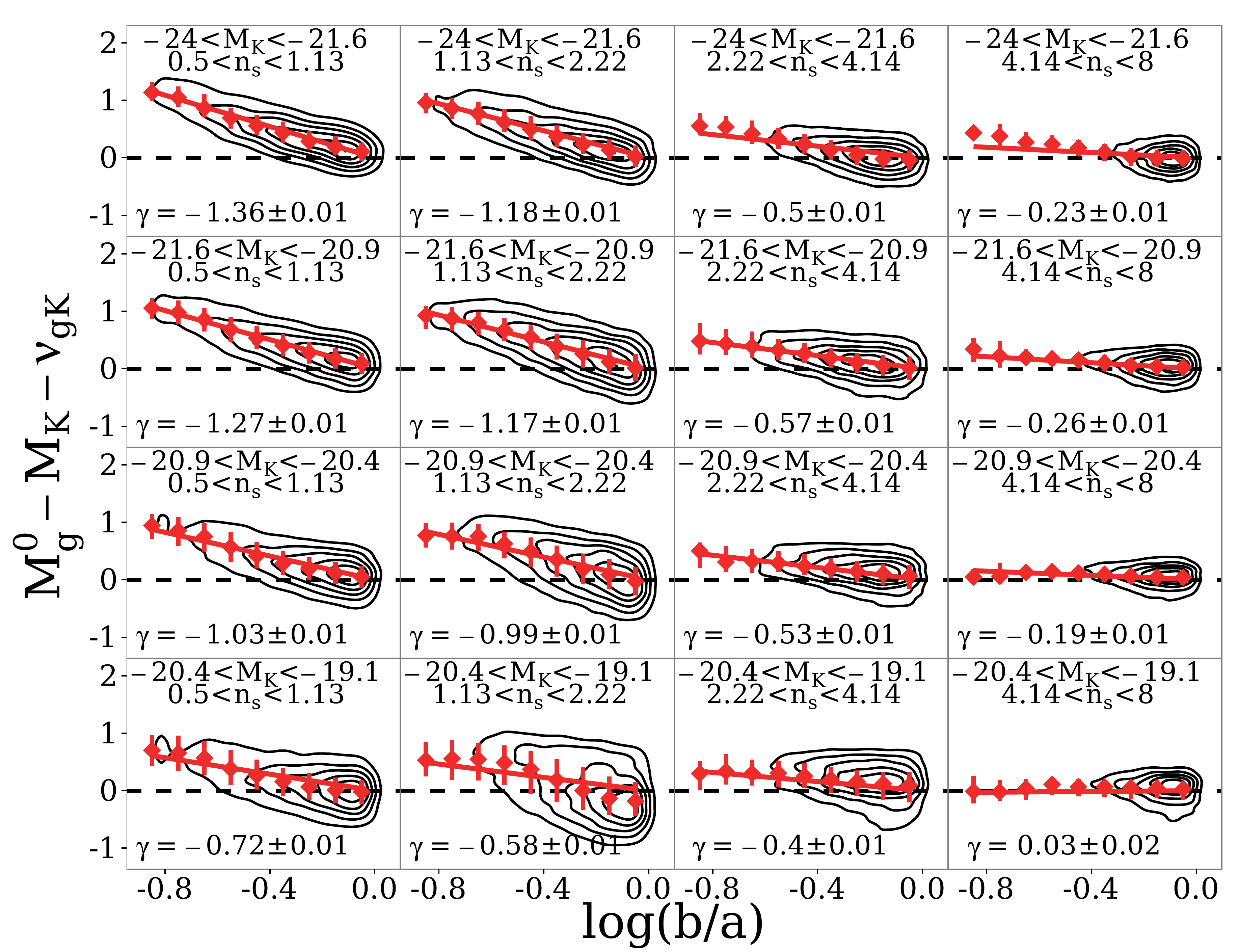} 
	\caption{Difference between M$_g$ - M$_K$ colour and the expected M$_g$ - M$_K$ colour of a face-on galaxy with the same M$_K$ and n$_s$ as a function of axis ratio. Results are binned by the quartiles of M$_K$ and n$_s$. Points show median values and interquartile ranges in log(b/a) bins. The best-fit slopes $\gamma$ for each panel are shown.} 
	\label{fig:Residual}
\end{figure*}

With a successful model for face-on $g-K$ and $r-K$ galaxy colours, we can examine how axis ratio affects the observed colours of samples of intrinsically identical galaxies. Fig \ref{fig:Residual} shows the residual between the M$_g$-M$_K$ colour and $\nu_{g-K}$ as a function of axis ratio binned into quartiles of M$_K$ and n$_s$. \textcolor{black}{While the strength of the trend varies across parameter space}, highly-inclined galaxies are systematically redder than identical face-on ones. Despite significant scatter, the trend is linear in log(b/a). We therefore express the attenuation as 

\begin{equation} \label{atten}
	A_{g/r - K} = \gamma_{g/r} \text{log}(b/a) \; \text{,}
\end{equation}
where $\gamma_g$ and $\gamma_r$ are slopes of a linear fit of $M_{g/r} - M_K - \nu_{g/r-K}$ to log($b/a$). Values for $\gamma_g$ are shown in Fig. \ref{fig:Residual} for each bin of M$_K$ and n$_s$. We notice that $\gamma_g$ depends on both M$_K$ and n$_s$. We obtain our best fit when we express $\gamma_g$ and $\gamma_r$ as

\begin{subequations}
\begin{align}
	\gamma_{\lambda} &= \gamma_{\lambda,0} + \gamma_{\lambda,K} (M_K + 20) + \gamma_{\lambda,n} n_s &\text{for n}_s<\text{3}\\
	\gamma_{\lambda} &= \gamma_{\lambda,0}' + \gamma_{\lambda,K}' (M_K + 20) &\text{for n}_s \geq \text{3}
\end{align} 
\label{gamma}
\end{subequations}

We find the best-fitting parameters for this model using a Monte Carlo Markov Chain to minimize

\begin{equation}
	\chi^2 = \sum \left[ \frac{M_{g/r} - M_K - \nu_{g/r-K} - A_{g/r-K}}{\sigma_{g/r-K}} \right]^2 \; \text{,}
\label{Chi}
\end{equation}
where $A_{g/r-K}$ is given by Eq. \ref{atten} and \ref{gamma} and $\sigma_{g/r-K}^2$ is the variance of $M_g/r - M_K - \nu_{g/r-K}$. We see no evidence of $\sigma_{\lambda-K}$ varying across parameter space so it is taken as a constant for both the n$_s<$3 and n$_s\geq$3 samples.
 
\begin{figure*}
	\includegraphics[width=0.95\textwidth]{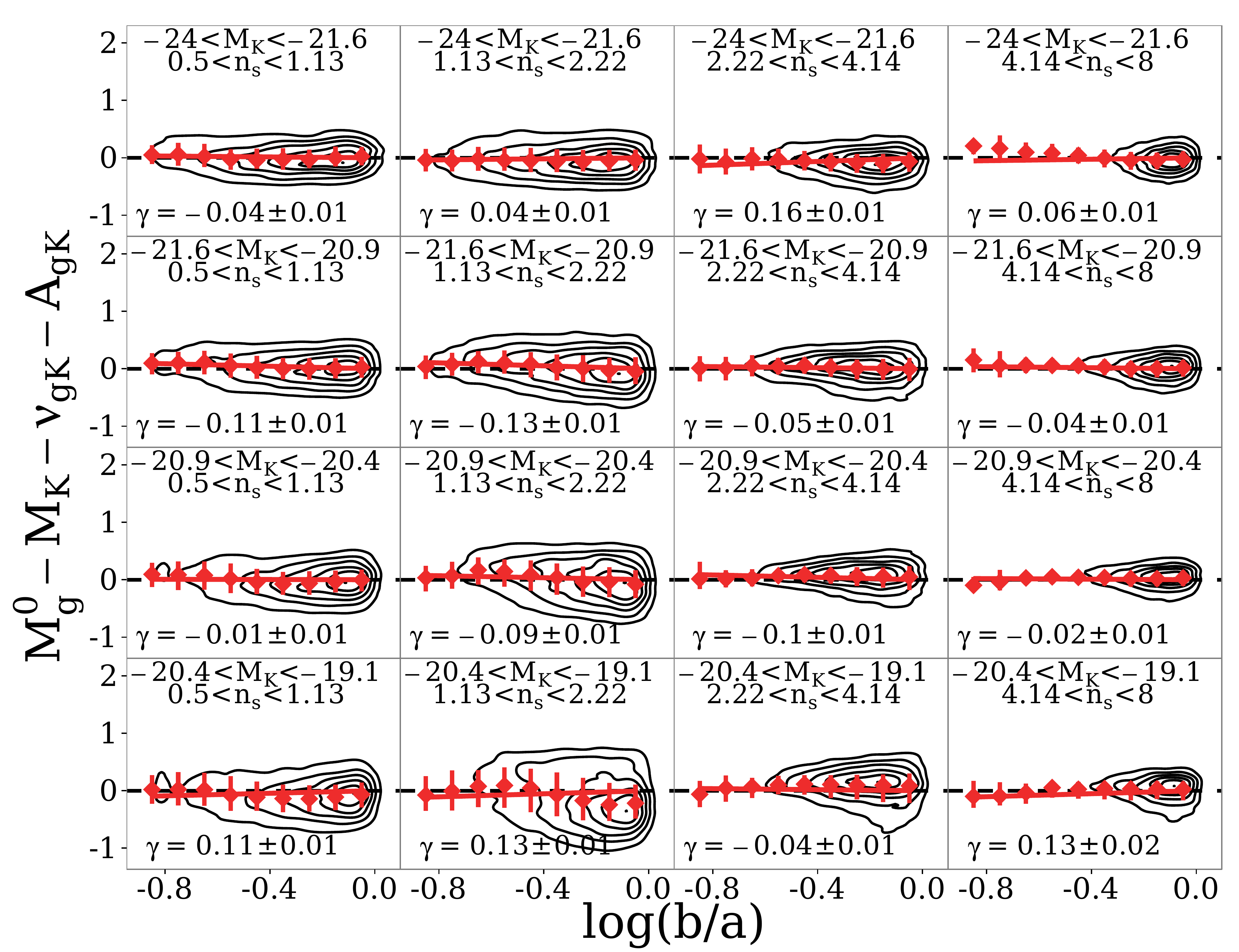} 
	\caption{Difference between \textit{corrected} M$_g$ - M$_K$ colour and the expected M$_g$ - M$_K$ colour of a face-on galaxy with the same M$_K$ and n$_s$ as a function of axis ratio. Results are binned by the quartiles of M$_K$ and n$_s$. Points show median values and interquartile ranges in log(b/a) bins.} 
	\label{fig:ResidualCorr}
\end{figure*}

With a model for the attenuation and best-fit parameters for that model, Fig. \ref{fig:ResidualCorr} shows the data from Fig. \ref{fig:Residual}
with $A_{g-K}$ subtracted. We see that the correction is successful at removing any trend between galaxy colour and axis ratio matching the M$_g$ - M$_K$ colours of edge-on galaxies to face-on ones. The correction is equally successful with M$_r$ - M$_K$ colours. 

With an $A_{g-K}$ and $A_{r-K}$ that successfully reverse the attenuating effect of inclination, we can finally obtain intrinsic $g-r$ colours for each galaxy:

\begin{equation}
	(M_g - M_r)^i = (M_g^0 - A_{g-K}) - (M_r^0 - A_{r-K}) \; \text{.}
\label{ColInt}
\end{equation}

Defining our populations using the local minima of the colour and sSFR distributions as described in \ref{Classification}, 46818 galaxies are classified as Red Misfits when we use uncorrected $(M_g - M_r)^0$. When using corrected $(M_g - M_r)^i$, the local minimum shifts bluewards by 0.02 mags and the Red Misfit population drops by 35 per cent, the rest being re-classified as Blue Actives. This correction is successful at mitigating the contamination of the Red Misfit population by highly-inclined Blue Actives.

\bsp	
\label{lastpage}
\end{document}